\documentclass[prb,twocolumn,aps,10pt,eqsecnum]{revtex4-1}

\usepackage{amsmath}
\usepackage{graphicx}
\usepackage{rotating}
\usepackage{multirow}
\usepackage{latexsym}
\usepackage{textcomp}
\usepackage{verbatim}
\usepackage{color}
\usepackage{pict2e}
\usepackage{bm}
\usepackage{subfigure}
\usepackage{everysel}
\usepackage{keyval}
\usepackage{ragged2e}
\usepackage{dsfont}
\usepackage{amssymb}
\usepackage{enumitem}
\usepackage{pstricks}
\usepackage{mathtools}
\usepackage{hyperref}

\newcommand{\beq}{\begin{equation}} 

\newcommand{\eeq}[1]{\label{#1} \end{equation}}

\newcommand{\cH}{\mathcal H}

\newcommand\bmb{\left( \begin{matrix}}
\newcommand\emb{\end{matrix} \right)}

\def\LSCO{La$_{2-x}$Sr$_x$CuO$_4$}
\def\LBCO{La$_{2-x}$Ba$_x$CuO$_4$}

\def\C60{A$_x$C$_{60}$}

\def\etal{{\it et al.}}

\def\HgCu3{HgCa$_2$Cu$_3$O$_{8+y}$}
\def\HgCu4{HgBa$_2$Ca$_3$Cu$_4$O$_{10+y}$}
\def\TlCu{Tl$_2$Ba$_2$CuO$_{6+\delta}$}
\def\TlCu3{Tl$_2$Ba$_2$Ca$_2$Cu$_3$O$_{10+y}$}
\def\TlCu4{Tl$_2$Ba$_2$Ca$_3$Cu$_4$O$_{12+y}$}

\def\BiCu3{Bi$_2$Sr$_2$Ca$_{2}$Cu$_3$O$_y$}

\def\8LSCO{La$_{1.88}$Sr$_{.12}$CuO$_4$}
\def\110LNSCO{La$_{1.5}$Nd$_{0.4}$Sr$_{0.1}$CuO$_{4}$}
\def\stage4LCO{La$_{2}$CuO$_{4+\delta}$}
\def\Y248{YBa$_2$Cu$_4$O$_8$}

\def\NbSe2{NbSe$_2$}
\def\TaSe2{TaSe$_2$}
\def\TiSe2{TiSe$_2$}
\def\NaCoOH2O{Na$_{0.3}$CoO$_{2y}$H$_2$O}
\def\MgB2{MgB${}_2$}

\def\URu2Si2{URu$_2$Si$_2$}

\def\Ba122{Ba(Fe$_{1-x}$Co$_x$)$_2$As$_2$}

\def\cJ{{\cal J}} 

\begin{document}

\title{Quasi One Dimensional Pair Density Wave Superconducting State}
\author{Rodrigo Soto-Garrido, Gil Young Cho, and Eduardo Fradkin}
\affiliation{Department of Physics and Institute for Condensed Matter Theory, University of Illinois, 1110 W. Green St., Urbana IL 61801-3080, U.S.A.}
\date{\today}

\begin{abstract}
We provide a quasi-one-dimensional model which can support a pair-density-wave (PDW) state, in which the superconducting (SC) order parameter 
modulates periodically in space, with gapless Bogoliubov quasiparticle excitations. 
The model consists of an array of strongly-interacting one-dimensional systems, 
where the one-dimensional systems are coupled to each other by local interactions and tunneling of the electrons and Cooper pairs between them. 
Within the interchain mean-field theory, we find several SC states from the model, including a conventional uniform SC state, 
PDW SC state, and  a coexisting phase of the uniform SC and PDW states. In this quasi-1D regime we can treat the strong correlation physics essentially exactly using bosonization methods and the crossover to the 2D system by means of interchain mean field theory. The resulting critical temperatures of the SC phases generically exhibit a power-law scaling with the coupling constants of the array, instead of the essential singularity found in weak-coupling BCS-type theories. 
Electronic excitations with an open Fermi surface, which emerge from the electronic Luttinger liquid 
systems below their crossover temperature to the Fermi liquid, are then coupled to the SC order parameters via the proximity effect. 
From the Fermi surface thus coupled to the SC order parameters, we calculate the quasiparticle spectrum in detail. 
We show that the quasiparticle spectrum can be fully gapped or nodal in the uniform SC phase and also in the coexisting phase of the uniform SC 
and PDW parameters. In the pure PDW state, the excitation spectrum has a  reconstructed Fermi surface in the form of Fermi pockets of Bogoliubov quasiparticles.
\end{abstract}

\maketitle

\section{Introduction}

The pair-density-wave (PDW) state is a superconducting (SC) state of matter in which the Cooper pairs have a finite momentum. 
Due to the finite momentum carried by the pair, the SC order parameter is modulated periodically in space. 
The PDW state has recently received attention because it can explain the layer decoupling observed in the cuprate \LBCO(LBCO), 
the original high $T_c$ superconductor,  at the $x=1/8$ anomaly.\cite{Berg-2007}
At this doping the $T_c$ of the three-dimensional material is suppressed to temperatures as low as $ 4$K, where the Meissner effect
first appears and the system is in a three-dimensional $d$-wave SC phase. 
In contrast, away from $x=1/8$, the SC $T_c$ in LBCO is about 35K. In spite of the low $T_c$ of the uniform $d$-wave SC state at $x=1/8$, 
in this doping regime the CuO planes are already superconducting for a range of temperatures well above $T_c$.\cite{Li-2007,tranquada-2008} 
With this in mind, Berg and collaborators\cite{Berg-2007,Berg-2009}  showed that this phenomenon can be explained if the CuO planes are in an 
inhomogeneous (`striped') SC state, the PDW state, in which charge, spin and SC orders are intertwined with each other.
In this state the SC order parameter oscillates along one direction in the CuO planes, and the average of the SC order parameter 
is zero in the CuO planes.  

The PDW state is similar to the traditional Larkin-Ovchinnikov (LO) state\cite{Larkin-1964}  where the Cooper pairs have a 
non-zero center of mass momentum, which in the LO proposal is due to the presence of an external Zeeman field, which breaks the time-reversal 
symmetry explicitly (for  a recent review of the LO state see Ref. [\onlinecite{Casalbuoni-2004}]). 
However, the occurrence of the PDW SC state does not necessarily require to have a system in which  time-reversal symmetry is explicitly broken, 
nor it does require time-reversal symmetry be spontaneously broken either. 
Since it was  proposed as a candidate competing state to the uniform $d$-wave SC order,\cite{Berg-2007, Berg-2009} the PDW state has been 
studied extensively. A Landau-Ginzburg (LG) theory of the PDW state provides a simple explanation of much of the observed phenomenology of 
{\LBCO},\cite{berg-2008a,Berg-2009,agterberg-2008} and of \LSCO { }in magnetic fields. An outgrowth of these phenomenological theories is a 
statistical mechanical description of the thermal melting of the PDW phase by proliferation of topological defects which yielded a rich phase 
diagram which includes, in addition to the PDW phase, a novel charge $4e$ SC state and a CDW phase.\cite{Berg-2010,barci-2011,fradkin-2014}
More recently, Agterberg and Garaud\cite{agterberg-2014} showed that it is possible to have a phase in which a uniform SC and PDW order parameters 
coexist in the presence of a magnetic field.

The microscopic underpinnings of the PDW state are presently not as well understood as the phenomenologies.  Nevertheless, it has been shown that this state can 
appear   in  different regimes of several models. In the weak coupling limit, such a state appears naturally in two dimensions (2D) 
inside an electronic spin-triplet nematic phase.\cite{Soto-Garrido-2014} 
Also at the mean field level, Lee \cite{lee-2014} found that it is possible to have a PDW state in his model of `Amperian' 
pairing\cite{sslee-2007} and that the PDW state can explain the pseudo-gap features found in the angle-resolved photoemission 
experiment.\cite{ARPES}  
In a series of papers,  Loder and collaborators\cite{Loder-2010,Loder-2011} found that a PDW superconducting state is preferred in a tight-binding model with strong attractive
interactions (although the critical value of the coupling constant above which the PDW is 
stable is presumably outside the range of validity of the weak coupling theory). Similarly, PDW states with broken time-reversal invariance and parity have been found recently\cite{Wang-2014,Wang-2015} in a `hot spot' model, which also requires a critical (and typically not small) value of a coupling constant.
On the other hand, in one-dimensional systems (1D)  the 
PDW state has been shown to describe the SC state  of the Kondo-Heisenberg chain,\cite{zachar-2001a,zachar-2001b,Berg-2010} and a 
phase of an extended Hubbard-Heisenberg model on a  two-leg ladder.\cite{jaefari-2012}
We showed recently that the PDW state appearing in these two 1D models is actually a topological SC 
which  supports Majorana zero modes localized at its boundaries.\cite{Cho-2014}

There has also been considerable recent  effort to determine if the PDW state occurs in simple models of  strongly correlated systems.
Variational Monte Carlo simulations of the $t-J$ and $t-t'-J$ model on the square lattice at zero magnetic field near doping $x=1/8$
found that the uniform $d$-wave SC state is slightly favored over the PDW state.\cite{Himeda-2002,Raczkowski-2007,Capello-2008,Yang-2008b}  
Corboz  and coworkers,\cite{corboz-2014} using infinite projected entangled pair-states\cite{Verstraete-2008} (iPEPS),  found 
strong evidence   in  the 2D $t-J$ model that the ground state energies of  the uniform $d$-wave state and the PDW state are numerically 
indistinguishable (within the error bars) over a broad range of dopings and parameters. 
This last result indicates that these strongly correlated systems do have a strong tendency to exhibit intertwined orders and that the PDW 
state occurs more broadly than was anticipated.\cite{fradkin-2014}

In this work, instead of following a conventional weak coupling approach to the PDW states, we will take an alternative path which has
the physics of strong correlations as its starting point. Rather than  starting from a true 2D system, we will consider a 
quasi-one dimensional model consisting of weakly coupled (each strongly-interacting) 1D systems. In the decoupled limit we can solve each 1D 
system non-perturbatively  using bosonization methods.\cite{FradkinFieldTheory,Gogolin-book,Giamarchi-book} We  will follow a dimensional 
crossover approach that has been used with considerable success by several authors.
\cite{Carlson-2000,emery-2000,granath-2001,vishwanath-2001,Carr-2002,essler-2002,Arrigoni-2004,jaefari-2010}
We will consider a generalization of the model used by Granath and coworkers\cite{granath-2001} in which there are two types of  1D subsystems: 
a set of doped two-leg ladders in the Luther-Emery (LE)  liquid regime (which has a single gapless charge sector and a gapped spin sector) 
and a set of 1D electronic Luttinger liquids (eLL) with both a gapless charge sector and a gapless spin sector. 
Although the  interactions between LE liquids and between LE and eLL liquids will be treated by the interchain mean field theory (MFT) 
(see, e.g. Carlson {\it et al.}\cite{Carlson-2000}), the intra LE and intra eLL interactions are treated essentially exactly using bosonization.
We will make the reasonable assumption that the interaction between the electronic Luttinger liquids leads to a crossover to a full 2D 
(anisotropic) Fermi liquid (see, e.g. Ref.[\onlinecite{granath-2001}]). In this fashion, this approach allows to access the strong coupling 
regime of a strongly correlated system using controlled approximations. 
In  this approximation the resulting superconducting $T_c$ is a power law in the interchain coupling and not exponentially
small as in the usual weak-coupling limit (such as BCS approach). 

The main departure of the system that we consider here  from previous studies of models on this type is that we will allow for the Josephson 
couplings between the LE liquids to have either positive or negative signs. A negative sign induces a $\pi$ phase shift 
between two neighboring LE liquids. It was shown by Berg {\it et al.}\cite{berg-2008a} that two superconductors that are proximity coupled to each other 
through a 1D weakly doped Hubbard model have a broad regime of parameters (in particular, doping)  in which the effective Josephson coupling is negative. 
Here, in order to incorporate this physics, we will introduce a set of Ising degrees of freedom mediating the interactions between the LE liquids which emulate different doping profiles 
of the electronic  Luttinger liquids. This feature will allow us to consider the interplay between uniform (s-wave or d-wave) superconductivity 
with PDW superconducting states and  coexistence phases, resulting in complex phase diagrams.

We note that inhomogeneous SC states such as the PDW are generally accompanied by a subsidiary 
charge-ordered state, a charge-density-wave (CDW). The period of the CDW is twice the period of the PDW or equal to the period of the PDW depending on whether this is a pure PDW state or if it is a state in which it coexists with an uniform SC state. The CDW order parameters which describe these states are composite of the PDW order parameters with or without the uniform SC order parameter. The general occurrence of charge-ordered states as subsidiary orders of an inhomogeneous SC state has been emphasized by several authors.\cite{berg-2008a,Berg-2009,agterberg-2008,fradkin-2014,lee-2014,Wang-2014} The same should hold in the case of the SC states that we study here.

The experimental consequences of the PDW states have been discussed extensively in the recent literature\cite{Berg-2007,agterberg-2008,Berg-2009,Berg-2009b,zelli-2011,zelli-2012,lee-2014,fradkin-2014} (including  papers by one of the authors) and we will not elaborate further on this questions here. Instead we will focus on microscopic mechanisms behind these inherently strongly interacting states.

The   paper is organized as follows. 
In section \ref{sec:Model} we define our model and summarize our notation for bosonization in 1D. 
In section \ref{sec:MFT} we develop the interchain MFT and discuss the results for the self consistent equations. 
In section \ref{sec:LLsystems} we study and discuss the quasiparticle spectrum of the phases, emergent from this quasi-1D system, 
for the various PDW and uniform SC states found in section \ref{sec:MFT}. 
In section \ref{sec:phases} we summarize other possible phases that could arise in this model using a qualitative scaling
dimensional analysis. We finish with our conclusions in section \ref{sec:conclusions}.

\section{The Model}
\label{sec:Model}

The quasi-1D model, schematically presented in the Introduction,  consists of two different types of 1D systems. 
One of them is a conventional 1D electronic Luttinger liquid (eLL)  in which both the spin and charge degrees of freedom are gapless. 
The other type, however, is a 1D system with a spin gap, i.e., it is a 1D Luther-Emery liquid (LE). The presence of the 
spin gap in the 1D system will bias the full array of 1D systems  to a strong tendency to a SC state. 

\subsection{1D Systems and  Bosonization}
\label{sec:1D}

Before we define in detail the quasi-1D model, 
we start with a short summary on those 1D liquids and their description using bosonization. This material is standard and can be found in 
several textbooks, e.g. Ref.[\onlinecite{FradkinFieldTheory}]. Here we will only give some salient results and set up our definitions 
(and conventions) that we  use in later sections.

 We start with a 1D eLL which has   a gapless charge sector and a gapless spin sector. 
The low-energy Hamiltonian is written in terms of the set of the bosonic fields, $\{ \phi_a, \theta_a \}$  where $a = c, s$ labels
the charge and spin sectors respectively. These fields 
satisfy canonical equal-time commutation relations 
\begin{equation}
[\phi_a (x'), \partial_x \theta_b (x)] = i \delta_{a, b} \delta(x'-x)
\end{equation}
The effective Hamiltonian for the eLL is
\begin{equation}
H_{eLL}[\phi_{a}, \theta_{a}] = \sum_{a = c,s} \frac{v_\alpha}{2} \left[ K_a (\partial_x \theta_a)^{2} + \frac{1}{K_a} (\partial_x \phi_a)^2 \right], 
\label{LL}
\end{equation}
in which $K_a$ (again with $a=c, s$) are the Luttinger parameters for the charge and spin sectors, 
and $v_a$ are the characteristic speeds for the charge and spin excitations of the 
liquid. The parameters $K_c$, $K_s$, $v_c$ and $v_s$  are determined by the microscopic details of the model. However, for a system with spin-rotational invariance, the resulting SU(2) symmetry restricts the value of   the the spin Luttinger parameter  to be $K_s=1$. 
In this continuum and low-energy limit,  
we can decompose  the electronic field operator in terms of two slowly varying components, with wave vectors near the two Fermi points $\pm k_F$
\begin{equation}
 	\frac{1}{\sqrt{a}}\psi_{\sigma}(x) \rightarrow R_{\sigma}(x)e^{ ik_{F}x} + L_{\sigma}(x)e^{-ik_{F}x}, 
\end{equation}
where $a$ is the ultraviolet cut-off (typically the lattice spacing), and where  $\sigma=\pm$ denotes the spin of the electron. Here the fermionic fields 
$R_\sigma (x)$ and $L_\sigma (x)$ are the right- and left-moving components of the electron field $\psi_{\sigma}$, 
which are slowly varying in space relative to the Fermi momentum $k_F$. 

The right- and left-moving fields can be written in terms of the bosonic charge fields $\phi_c$ and $\theta_c$, and the spin fields $\phi_s$ and $\theta_s$, as follows
\begin{align}
 R_{\sigma}&=\frac{F_{\sigma}}{\sqrt{2\pi a}}e^{i\sqrt{\pi/2}(\theta_{c}+\sigma\theta_{s}+\phi_{c}+\sigma\phi_{s})},\nonumber\\ 
 L_{\sigma}&=\frac{F_{\sigma}}{\sqrt{2\pi a}}e^{i\sqrt{\pi/2}(\theta_{c}+\sigma\theta_{s}-\phi_{c}-\sigma\phi_{s})}.
\end{align}
The anticommuting Klein factors, $F_{\sigma}$, ensure the fermionic statistics between the right and left moving fermions $ R_\sigma$ and  $L_\sigma$.

Next we consider a spin-gapped Luttinger liquid, or LE liquid. At energy scales below the spin gap $\Delta_s$, 
the spin sector can be ignored. Hence, the low-energy Hamiltonian contains only the charge fields $\phi_c$ and $\theta_c$, and it is given by
\begin{equation}
H_{LE} [\phi_c, \theta_c] = \frac{v_c}{2} \left[ K_c (\partial_x \theta_c)^{2} + \frac{1}{K_c} (\partial_x \phi_c)^2 \right]. 
\label{LE}
\end{equation}
Since the spin sector has been effectively projected-out, we will keep only the charge sector of the LE liquid and drop its $c$ label.

In the LE liquid, all interactions represented by operators that are not spin singlets are irrelevant 
(and, in fact, with effective scaling dimension infinite). This fact strongly restricts the types of interactions between LE systems and 
eLL systems. In this case the only fermion bilinears that need to be considered in  the LE liquid are the order parameter of the charge-density-wave  with momentum $2k_F$ (CDW) 
\begin{equation}
O_{\text{CDW}}(x) \sim e^{-2ik_F x} R^{\dagger}_{\sigma}(x) L_{\sigma}(x) + h.c.,
\end{equation} 
and the (Cooper) pair field spin singlet superconducting order parameter

\begin{equation}
\Delta (x) \sim R_\alpha(x) (i\sigma^y)^{\alpha\beta} L_{\beta}(x)  + (R \leftrightarrow L). 
\end{equation}
Hence the coupling to the LE liquids in the quasi-1D model should involve only  the two operators listed above. 
We note that the suppression of the spin operator and the power-law correlation for the SC operator make the LE liquids the natural 
building blocks for the quasi-1D SC state. 
In contrast, the eLL has other observables that need to be considered, including a spin triplet pair field, the $2k_F$ spin-density-wave (SDW) `N\'eel' order parameter,  the right and left moving spin current operators, and, in tunneling processes, the electron operators.

\subsection{Quasi-1D Model}
\label{sec:quasi-1D}

\begin{figure}
\begin{center}
\includegraphics[width=\columnwidth]{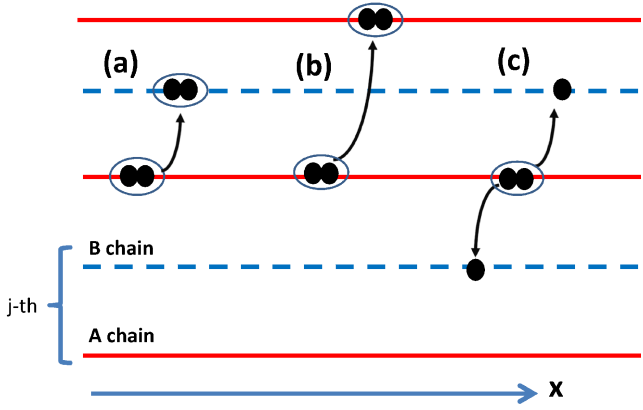}
\caption{
(color online) The array of LE systems and eLL systems. The $A$-type LE systems are represented by the solid (red) line. 
The $B$-type eLL systems are represented by the dashed (blue) line. Each unit cell consists of one $A$-type and one $B$-type systems. 
Here the black filled dots represent electrons. (a) The conventional Josephson coupling ${\cal J}_{AB}$ in Eq.\eqref{Hint} (b) The 
conventional Josephson coupling ${\cal J}_{AA}$ in Eq.\eqref{Hint} (c) Splitting a Cooper pair in $A$ system into the neighboring $B$ 
systems ${\cal J}_{AB}^{\prime}$ in Eq.\eqref{Hint}.  
 \label{Fig_Chains}
}
\end{center}
\end{figure}

Given a set of   independent 1D LE and eLL systems that were described above, we now define and discuss the full quasi-1D model. 
The model consists of an array of 1D systems shown  in 
Fig. \ref{Fig_Chains}. Each {\it unit cell} of the array consists of one LE system, labeled by $A$, and one eLL system, labeled by $B$. 
Hence we introduce the bosonic fields $\{\phi_{n, A}, \theta_{n,A} \}$ representing the charge fields in the LE chain of the $n$-th unit cell 
and also $\{ \phi_{n,B,a}, \theta_{n,B,a} \}$ (with $a=c,s$) representing the charge and spin fields in the eLL chain of the $n$-th 
unit cell. Furthermore we assume that the filling of the type B system (an eLL chain) is close to half filling, i.e., $k^{(B)}_F\approx\pi/2$ 
and $K^{(B)}_c\approx 1/2$. Also the spin rotational symmetry in the B systems is assumed to be unbroken and thus $K^{(B)}_s = 1$. 
We further assume that the Fermi momenta of the systems $A$ and $B$ are incommensurate to each other. 

In the limit in which the LE systems and the eLL systems are decoupled from each other, the effective Hamiltonian of the array is described by the 
sum of Eq.\eqref{LL} and Eq.\eqref{LE} for each system, and has the form
\begin{equation} 
H_{0} = \sum_{n \in {\mathbb Z}} \Big( H_{LE} [\phi_{n,A}, \theta_{n,A}] + H_{LL} [\phi_{n,B, a}, \theta_{n,B,a}] \Big). 
\end{equation}
This decoupled limit  is an unstable fixed point and the system will eventually flow to the quasi-1D or 2D fixed points under the introduction 
of the coupling between the 1D systems. We will show that the PDW state, as a quasi-1D fixed point, will emerge from certain  couplings.

Following the work of Granath {\it{et al.}}, \cite{granath-2001} we write down the possibly relevant local interactions terms between the 1D systems. They are given by
\begin{align}
 H^{\prime} =&\sum_{n}\int dx \Big\{
-t_{BB}\sum_{\sigma}[\psi_{B,j,\sigma}^{\dagger}\psi_{B,j+1,\sigma}+ {\rm h. c. }]\nonumber\\
&+ J_{BB}{\bm S}_{B,j}\cdot {\bm S}_{B,j+1} \nonumber\\
&-{\cal J}_{AA,j} [\Delta^{\dagger}_{A,j+1}\Delta_{A,j} + {\rm h.c.}] \nonumber \\
&-{\cal J}_{AB}[\Delta^{\dagger}_{B,j}\Delta_{A,j}+\Delta^{\dagger}_{B,j}\Delta_{A,j+1}+{\rm h. c.}]\nonumber \\
&+{\cal J}_{AB}^{\prime}[\Delta_{A,j}^{\dagger}(\psi_{B,j,\uparrow}\psi_{B,j-1,\downarrow}+
\psi_{B,j-1,\uparrow}\psi_{B,j,\downarrow})+{\rm h. c.}]
\Big\}
\label{Hint}
\end{align}
To simplify the analysis, in this paper we will not consider the possible existence of spin-ordered phases (i.e. spin stripes or SDWs) although these are clearly seen in {\LBCO} which is the material where the PDW state is most clearly seen. We are mainly concerned about the SC states in which the spins 
do not play much role, and thus we  ignore for now the antiferromagnetic interactions in the discussion.
The antiferromagnetic coupling between the eLL chains can also be included in a relatively straightforward extension of the present work. 
Following Ref. [\onlinecite{granath-2001}] we have ignored the possible CDW couplings between chains. In general the scaling
dimensions of the CDW operators become less relevant in the presence of forward scattering interactions between the chains, \cite{emery-2000} so
they can be neglected.  If the interchain CDW couplings were to become relevant we would have bidirectional charge order. In this paper we are only exploring states with unidirectional charge and superconducting order.

In Eq.\eqref{Hint}, the operator $\Delta_{A, j}(x) \sim \psi_{A,j, \alpha}(x) (i\sigma^y)^{\alpha\beta}\psi_{A,j, \beta}(x)$ represents the density of the 
spin-singlet Cooper pair of the system $A$, and $\Delta_{B,j}(x)$ is that of the system $B$.    
The effective coupling constants ${\cal J}_{AB}$ and ${\cal J}_{AA,j}$ are  the conventional Josephson coupling between the two neighboring $A$ systems, 
representing the hopping process of the Cooper pairs   (see the (a) and (b) in Fig.\ref{Fig_Chains}). 
On the other hand, the local term ${\cal J}_{AB}'$ represents the breaking of a Cooper pair in an $A$ system which puts the two single electrons 
into the nearest neighbor $B$ systems (see the (c) in Fig.\ref{Fig_Chains} for a diagram of the process).  

In the Hamiltonian $H'$, Eq.\eqref{Hint}, the most relevant term is the electron tunneling term, with coupling strength $t_{BB}$, 
between two nearest-neighbor $B$ systems. Under this perturbation, the decoupled $B$ systems flow to the 2D Fermi liquid fixed point, 
which, in turn,  becomes  coupled to the superconducting state emergent from $A$ systems.\cite{granath-2001} Due to this dimensional crossover of the $B$ systems 
it is difficult to apply the conventional interchain MFT to analyze Eq.\eqref{Hint}. In order to make progress, we ignore at first  the $B$ 
systems, as the first order approximation to the problem, and perform the interchain MFT only with the $A$ systems, which embodies the strong-coupling 
nature of the superconductivity emergent in the quasi-1D models. We should stress that in the $A$ systems, there are no electron-like quasiparticles due the existence 
of the spin gap which leads to a fully gapped  2D SC phases when the coupling between the $A$ systems are turned on. 
Technically speaking, we solve first for an array of $B$ (eLL) systems  coupled by $t_{BB}$ and for an array of $A$ systems coupled only 
by ${\cal J}_{AA,j}$ in Eq.\eqref{Hint}, and take ${\cal J}_{AB}'$ and ${\cal J}_{AB}$ as 
 perturbations. At this level of the approximation, the emergent SC state is determined by the Josephson coupling 
${\cal J}_{AA,j}$ and the subsequent SC state of the full system follows by proximity effect between the $A$ and the $B$ subsystems. 
This was the strategy used by Granth {\it et al.}.\cite{granath-2001}  The main difference between this work and that of Granath and coworkers is the inclusion of an additional, Ising-like, degree of freedom in the coupling between the $A$ systems, as we already discussed in the Introduction. 

\subsection{Coupled LE Systems}
\label{sec:LEsystems}

It is clear that ${\cal J}_{AA}$ will determine the nature of the emergent 2D SC state from the quasi-1D model. 
More precisely, the spatial pattern of ${\cal J}_{AA,j}$ determines that of the Cooper pair. For example, 
if the Josephson coupling ${\cal J}_{AA,j}$ in Eq.\eqref{Hint} is uniform and positive, 
it is clear that the uniform spin-singlet SC will emerge. However, in the strongly-correlated quasi-1D system, 
the Josephson coupling ${\cal J}_{AA,j}$ may not be always uniform and positive. 
In Ref. [\onlinecite{Berg-2009}], the Josephson coupling between two systems with an intermediate chain 
(which is close to the insulator phase) has been calculated by a numerical density matrix renormalization group (DMRG) 
method and it was found that it can be negative, i.e., forming a $\pi$-Josephson junction between the two $A$ systems. From this, 
it is not difficult to imagine that there might be more complicated patterns, depending on the microscopic details, 
than the uniform $\pi$-Josephson junction. 

To reflect this physics and to consider a broader possible patterns of the Josephson coupling ${\cal J}_{AA,j}$, we introduce a {\it phenomenological} Ising degree of freedom 
$\sigma_{j}$ which can change the magnitude and possibly the sign of effective Josephson coupling. This Ising degree of freedom can be regarded as a local change in the doping of the intervening $B$ system between two neighboring $A$ systems. In this sense the Ising degree of freedom should be regarded as reflecting the tendency to frustrated phase separation of a doped strongly correlated system.\cite{emery-1993,carlson-1998} 
To this effect, we consider the following interaction between the Ising degrees of freedom and the LE liquid 
\begin{align}
 \cH_{\text{int}}=&-{\cal J}_{AA} \sum_{i}\cos[\sqrt{2\pi}(\theta_{A, i}-\theta_{A, i+1})]\nonumber\\
 &- {\cal J}^{\prime}_{AA} \sum_{i}\sigma_{i}\sigma_{i+1}\cos[\sqrt{2\pi}(\theta_{A, i}-\theta_{A, i+1})] \nonumber\\ 
 &+ \cH_{\text{Ising}}[\sigma_i], 
\label{Int}
\end{align}
in which we write ${\cal J}_{AA,i}$ in Eq.\eqref{Hint} as ${\cal J}_{AA,i}= {\cal J}_{AA} - {\cal J}'_{AA}\sigma_i \sigma_{i+1}$. The factor $\sim \cos[\sqrt{2\pi}(\theta_i-\theta_{i+1})]$ is the Josephson coupling between the LE systems because of 
$\Delta(x) \sim e^{i\sqrt{2\pi} \theta_{A,i}}$.  

In Eq.\eqref{Int}, the Ising interaction Hamiltonian $\cH_{\text{Ising}}[\sigma_i]$ is assumed to have several phases depending on the 
parameters in $\cH_{\text{Ising}}$ and temperature, e.g. paramagnetic phase $\langle \sigma_i \rangle =0$, and various symmetry-broken phases. 
In this paper, we further assume that the Ising variable $\sigma_i$ orders at a much higher temperature (or energy scale) than the spin gap 
$\Delta_s$ in the LE liquid. Hence, we ignore any correction to the Ising variable due to the fluctuations of the SC states emergent from LE 
liquids. To simplify the analysis we have assumed that the Ising variables are constant along the direction of the 1D systems and are classical (i.e. we did not include a transverse field term). The first assumption is not a problem since we will do mean field theory assuming that the resulting modulation (if any) is unidirectional. More microscopically we will need to assume that the Ising model has frustrated nearest and next nearest neighbor interactions along one direction only. This is the so-called anisotropic next-nearest-neighbor Ising (ANNNI) model which is well known to have a host of modulated phases.\cite{Fisher-1980} Similar physics, with a rich structure of periodic and quasi-periodic states,  is obtained from the Coulomb-frustrated phase separation mechanism.\cite{carlson-1998,Low-1994} 

In what follows we will not specify the form of $\cH_{\text{Ising}}$ and  assume that its ground state is encoded in a specific pattern of order for the Ising variables. In this picture an inhomogeneous charge-ordered state occurs first (and hence has a higher critical temperature) and this pattern causes the effective Josephson couplings to have  an ``antiferromagnetic'' sign (i.e. $\pi$ junctions).\cite{berg-2008a,Berg-2009} Nevertheless, as noted in Ref. [\onlinecite{Berg-2009}], once the PDW state sets in there is always a (subdominant) CDW order state with twice the ordering wave-vector as that of the PDW.

The symmetry breaking patterns that  we to study are: i) Uniform configuration $\langle \sigma_i \rangle = \pm$, $\forall i$, 
ii) Staggered configuration $\langle \sigma_i \rangle = (-1)^i$, and
iii) Period 4 configurations (which will become clear soon below \ref{sec:p4}). Thus, when the Ising variables order and 
spontaneously break the translational symmetry, the effective Josephson coupling between the different A systems will be modulated too. 
For concreteness, throughout this work, we will take ${\cal J}_{AA}$ and ${\cal J}_{AA}^{\prime}$ to be positive. This condition is not necessary 
and the following arguments can be easily extended to the other signs of ${\cal J}_{AA}$ and ${\cal J}_{AA}^{\prime}$. 

We will start by analyzing the ground state (or the mean field (MF) state) of the LE systems coupled to Ising variables. 
We will do this for different configurations of the Ising variables and see what are the possible phases that arise in the system of coupled 
LE liquids.

\subsubsection{Ising Paramagnetic Configuration}

Before proceeding to the symmetry-broken phases of the Ising variable, we first briefly comment on the case with the paramagnetic phase 
of the Ising variable $\sigma_i$. In the Ising paramagnetic phase, 
we first note that ${\cal J}'_{AA}\sigma_i \sigma_{i+1} \cos[\sqrt{2\pi}(\theta_i-\theta_{i+1})]$ is effectively zero at the level of mean field theory and can be ignored. Thus Eq.\eqref{Int} will become at the low energy  
\begin{align}
 \cH_{\text{int}} \rightarrow -{\cal J}_{AA}\sum_{i}\cos[\sqrt{2\pi}(\theta_{A, i}-\theta_{A, i+1})]+ \cdots 
\label{int_pm}
\end{align}
in which $\cdots$ are the terms generated by integrating the fluctuations of the Ising variables in the paramagnetic phase, 
e.g., $\sim \cos[2\sqrt{2\pi}(\theta_{A, i}-\theta_{A, i+1})]$, 
which is strictly less relevant than $-{\cal J}_{AA} \cos[\sqrt{2\pi}(\theta_{A,i}-\theta_{A, i+1})]$ appearing in $\cH_{\text{int}}$. 
It is well-known that Eq.\eqref{int_pm}  induces an uniform 2D superconducting state.\cite{Carlson-2000,Arrigoni-2004}

\subsubsection{Uniform Ising Configuration}
\label{sec:p0}

We now analyze the simplest case with $\langle \sigma_i \rangle\neq0$, where all the $\sigma_i$ have the same value, $\sigma_i=\sigma=\pm$. 
In this case $ H_{\text{int}}$ is just given by
\begin{align}
 \cH_{\text{int}}=&-({\cal J}_{AA}+ {\cal J}^{\prime}_{AA})\sum_i \cos[\sqrt{2\pi}(\theta_{A,i}-\theta_{A, i+1})]\nonumber\\
 \equiv&-{\cal J}_T\sum_i \cos[\sqrt{2\pi}(\theta_{A,i}-\theta_{A, i+1})]
\end{align}
 The system of coupled LE systems can be treated in interchain MFT, where all the systems are in phase, since ${\cal J}_T>0$. In this case we just have a 
 uniform SC state in the direction perpendicular to the systems $\Delta_{j}=\Delta$, where $\Delta$ includes the
 spin gap and the MFT value for $\langle\cos\sqrt{2\pi}\theta_{A,i} \rangle$. We will show in the following section how to compute the 
 value $\langle\cos\sqrt{2\pi}\theta_{A,i}\rangle$. Thus, this is the same phase as in the Ising paramagnetic case but with a larger value of the effective Josephson coupling.
 
\subsubsection{Staggered (Period 2) Ising Configuration}
\label{sec:p2}

Let us now consider $\sigma_i=(-1)^i$. In this case $\mathcal{H}_{\text{int}}$ is given by
\begin{align}
 \cH_{\text{int}}&=-({\cal J}_{AA}- {\cal J}^{\prime}_{AA})\sum_i \cos[\sqrt{2\pi}(\theta_{A,i}-\theta_{A,i+1})]\nonumber\\
 \equiv&-\delta {\cal J}\sum_i \cos[\sqrt{2\pi}(\theta_{A,i}-\theta_{A,i+1})]
 \label{HintLE}
\end{align} 
Again, the system of coupled LE systems can be treated in interchain MFT. However, we need to be careful about the sign of $\delta{\cal J}$. If 
$\delta {\cal J}>0$ the SC order parameter in all the systems are in phase. It is important to emphasize that although all the systems are in phase 
as in the uniform Ising configuration, the expectation value $\langle\cos\sqrt{2\pi}\theta_{A,i}\rangle$ is different in both cases, 
since as we will see below, it explicitly depends on the coupling between the systems, in this case ${\cal J}_T$ or $\delta {\cal J}$. 
On the other hand, if $\delta {\cal J}<0$ the phase of SC order parameter has a  $\pi$ phase shift between nearest neighbors. 
In the former case we just have a uniform superconducting state in the direction perpendicular to the systems, while in the second case we have a 
PDW state $\Delta_{A, j}\sim (-1)^{j}$. There is a direct transition from the uniform SC state to the PDW SC state at  
${\cal J}_{AA}/{\cal J}^{\prime}_{AA}=1$. In this simple period 2  Ising configuration there is no 
room for coexistence between the uniform SC and the PDW state. 

\subsubsection{Longer Period  Ising Configurations}
\label{sec:p4}

We can generalize the phases  obtained with period 2 Ising configurations to cases with longer periods of the Ising variables. For instance for a period 4 of the Ising variables, 
\begin{equation}
\cdots, \uparrow,  \uparrow, \downarrow, \downarrow,  \cdots, 
\end{equation}
the effective Josephson couplings will have a period 2 modulation. In this case we will find either an uniform SC state or a period 4 PDW SC state, but no coexistence phase.

However, we will see that for  Ising configurations with period $n$, with $n>2$, we 
can have a richer phase diagram,  including a coexistence phase if $n\geq 3$. For example, for a   period 3 structure of the Ising variables, the allowed SC  state is a coexistence phase, whereas for period 8 with the following  spatial pattern of the Ising degrees of freedom  
\begin{equation}
\cdots \downarrow, \uparrow, \uparrow, \uparrow, \uparrow, \downarrow, \downarrow, \downarrow, \downarrow, \uparrow, \uparrow, \cdots, 
\end{equation}
we will find either a coexistence phase with period 4 or a PDW SC with period 8. It is straightforward to generalize this to more intricate configurations of the Ising variables.

\section{Interchain MFT on the LE Systems} 
\label{sec:MFT}

Keeping the quasi-1D model of the previous section in mind, we now solve the coupled LE system problem using the interchain MFT. 
In this section, we generalize the works of Lukyanov and Zamolodchikov,\cite{Lukyanov-1997} and Carr and Tsvelik\cite{Carr-2002} 
to the patterns of the Josephson coupling between the LE systems emergent from various symmetry-breaking phases of the Ising variables. 

\subsection{Uniform SC and Period 2 PDW SC Phases}
\label{sec:unifstag}

We first review the uniform configuration of the Ising variable (and also the paramagnetic phase of the Ising variable), 
in which the SC operator will develop the same expectation value for all the LE systems\cite{Lukyanov-1997,Carr-2002}. 
For the staggered (period 2) Ising configuration, there are two phases, depending on the sign of $\delta {\cal J}$, a uniform SC state and a PDW state. 
We will solve the self-consistency equations for the both phases, the uniform SC state and a PDW state. Although 
the equations have the same form, they correspond to different phases. The case of a period 4 Ising configuration of the form $\uparrow,  \uparrow, \downarrow, \downarrow$ can be treated in the same manner. The only difference is that the two phases will be an uniform SC state or a period 4 PDW SC state. Here we will focus in the simpler period 2 case.

\subsubsection{Uniform SC Phase}

In the uniform configuration of the Ising variable, the effective Josephson interaction between neighboring $A$ subsystems (the LE liquids)  is
 \begin{align}
 \cH_{\text{int}}= -{\cal J}_T\sum_i \cos[\sqrt{2\pi}(\theta_{A, i}-\theta_{A, i+1})]
\label{uniform}
\end{align}
To perform the interchain MFT, we consider only the terms involving the $i$-th type-$A$ system among $\cH_{\text{int}}$. 
Using standard interchain MFT \cite{Carlson-2000,Carr-2002,Arrigoni-2004} we can approximate eq. \eqref{uniform} by:
\begin{equation}
 \cH_{\text{int}}=- 2\mu\int d^2x\cos(\sqrt{2\pi} \theta_{A,i}),  
\label{HintSG1}
 \end{equation}
with 
$2\mu={\cal J}_T[\langle\cos(\sqrt{2\pi}\theta_{i+1})\rangle+\langle\cos(\sqrt{2\pi}\theta_{i-1})\rangle]$. 
The self-consistency of the MFT then requires that
\begin{equation} 
\langle \cos(\sqrt{2\pi} \theta_{A,i})\rangle = 
\frac{\mu}{{\cal J}_T }. 
\label{self}
\end{equation}
Following Refs. [\onlinecite{Lukyanov-1997},\onlinecite{Carr-2002}] the self-consistency equation can be solved from the following two expressions:
\begin{align}
 &\langle \cos (\sqrt{2\pi}\theta_{A,i})  \rangle
= \frac{(1+\xi) \pi \Gamma(1-d/2)}{16 \sin \pi\xi\ \Gamma(d/2)}\times \nonumber\\ 
&\quad \left(\frac{\Gamma(\frac{1}{2}+\frac{\xi}{2})\Gamma(1-\frac{\xi}{2})}
{4\sqrt{\pi}}\right)^{(d-2)}\left( 2\sin \frac{\pi\xi}{2} \right)^d 
M^d
\label{MFcos}
\end{align}
where 
$M$, the soliton mass in the $1+1$-dimensional sine-Gordon model, is related to $\mu$ by
\begin{equation}
 \mu = \frac{\Gamma(d/2)}{\pi\Gamma(1-d/2)} \left(
\frac{2\Gamma(\xi/2)}{\sqrt{\pi}\Gamma(\frac{1}{2}+\frac{\xi}{2})}
\right)^{d-2} 
M^{2-d}
\label{mutoM}
\end{equation}
In these equations $d=1/(2K_c)$ is the scaling dimension of the vertex operator $e^{i\sqrt{2\pi} \theta_{A,i}}$ and $\xi=\tfrac{1}{2-d}$. Using equations Eq.\eqref{MFcos} 
and  Eq.\eqref{mutoM}, we can compute explicitly the value of $\langle\cos(\sqrt{2\pi}\theta_{A,i})\rangle$ for a given value of ${\cal J}_T$ 
and $K_c$. This completely determines, at least at the mean field level, the solution of the coupled LE systems \cite{Carlson-2000,Arrigoni-2004}.

\subsubsection{Period 2 PDW SC Phase } 
\label{sec:staggered}

In the staggered configuration of the Ising variable, the  interaction term between the $A$ systems  is 
 \begin{align}
 \cH_{\text{int}}= -\delta {\cal J} \sum_i \cos[\sqrt{2\pi}(\theta_{A, i}-\theta_{A, i+1})], 
\end{align}
which is identical to that of the uniform configuration case, Eq.\eqref{uniform}, if $\delta {\cal J}>0$. Hence if $\delta {\cal J} >0$, 
we can simply replace ${\cal J}_T$ by $\delta {\cal J}$ to find the MF solution. This will give  a uniform SC state. 
  
If $\delta {\cal J} <0$, then we can perform a transformation on the even sites, $\sqrt{2\pi}\theta_{A, 2i}\to\sqrt{2\pi}\theta_{A, 2i}+\pi$, 
effectively changing the sign of $\delta J$ and coming back to the first case. 
Though the form of the equation is identical to that of the uniform SC state, it is important to remember 
that the MF solution doubles the unit cell, due to the transformation $\sqrt{2\pi}\theta_{A, 2i}\to\sqrt{2\pi}\theta_{A, 2i}+\pi$ 
acting only on the even sites. Thus, SC order parameter oscillates in space 
\begin{equation}
\Delta_{j}(x) \sim (-1)^j \langle \cos(\sqrt{2\pi}\theta_{A}) \rangle, 
\end{equation} 
corresponding to a period-2 PDW SC state. 

Before moving onto the coexistence phase in the next section, let us mention what is the dependence of $T_c$ with $\delta\cJ$ (or $\cJ_T$, 
depending on the Ising configuration). We can think of $2\mu$ in eq. \eqref{HintSG1} effectively as an external field due to the mean field
value of $m_j=\langle\cos(\sqrt{2\pi}\theta_{j})\rangle$ in the nearest neighbor systems. We can write then
\begin{equation} 
H_j = H^{(0)}_j - h_j \int dx \cos(\sqrt{2\pi} \theta_j)
\end{equation} 
in which $h_j=\cJ(m_{j+1}+m_{j-1})$ and $H^{(0)}_j$ is the conventional kinetic term for the Luther-Emery liquid. As we saw above, for 
the uniform or staggered configuration the value of $m_j$ is the same in all the systems, or effectively the same for 
$\delta {\cal J} <0$ since we can perform a transformation on the even sites $\sqrt{2\pi}\theta_{A, 2i}\to\sqrt{2\pi}\theta_{A, 2i}+\pi$.

In summary, we can write just $m=m_j=\langle\cos(\sqrt{2\pi}\theta_{j})\rangle$ and $h=h_j=2\cJ m$ (where $\cJ=\delta\cJ$ or $\cJ_T$ 
depending on the case).
For $h\to0$ we have that self-consistency implies 
\begin{equation}
m = \chi_{SC} h  = 2\cJ \chi_{SC} m, 
\end{equation}
which has the trivial solution $m=0$ or a non-trivial solution $m\neq0$ if $2\cJ \chi_{SC} = 1$ (which determines the critical temperature). 
Using that for a Lutter-Emery liquid 
\begin{equation}
\chi_{SC}(T) \sim \frac{\Delta_{s}}{T^{2- 1/K_c}}, 
\end{equation}
we have that:
\begin{equation}
 T_c\sim \Delta_s\cJ^\alpha
 \label{Tcpure}
\end{equation}
where the exponent is $\alpha=\displaystyle\frac{1}{2- 1/K_c}$. Although the resulting $T_c$ is small when $\cJ$ is small, what is important is that  is only power-law small, instead of exponentially small as in the BCS case.

\subsection{Uniform SC and Period 4 PDW SC state coexistence phase} 
\label{sec:coex}

Now we consider the period 8 states of the Ising variables $\sigma_i=(-1)^{\lfloor i/4\rfloor}$. Then the Josephson coupling also modulates in
space with period 4, and thus we need to solve four coupled self-consistent equations in MFT. The effective MF Hamiltonian for each $A$ system 
is given by\cite{Carr-2002}
\begin{equation}
 H^{(i)}_{\text{int}}=- 2\mu_i\int d^2x\cos(\sqrt{2\pi}\theta_{A,i})
\end{equation}
with  
\begin{equation}
2\mu_i=[{\cal J}_{i}\langle\cos(\sqrt{2\pi}\theta_{A, i+1})\rangle+{\cal J}_{i-1}\langle\cos(\sqrt{2\pi}\theta_{A, i-1})\rangle]
\end{equation}
 where ${\cal J}_i = {\cal J}_{AA}- {\cal J}^{\prime}_{AA}\sigma_i \sigma_{i+1}$ in which $\sigma_i$ is in the period 4 structure.
 
Using $\cJ_T=\cJ_{AA}+ \cJ^{\prime}_{AA}$ and $\delta \cJ=\cJ_{AA}- \cJ^{\prime}_{AA}$ and defining 
$m_i= \langle \cos(\sqrt{2\pi} \theta_{A,i}  \rangle$, it is clear that we need to solve only for the four systems $i=0, 1, 2, 3$ 
in this MFT by assuming that the MF solution does not break the translational symmetry $i \sim i+4$ of the pattern of the Josephson coupling.

Upon implementing the MFT analysis from the previous section we have the following set of coupled equations:
\begin{align}
 m_0&=f(d)\left(\frac{ m_3\delta {\cal J}+m_1{\cal J}_T }{2}\right)^{d/(2-d)},\nonumber\\
 m_1&=f(d)\left(\frac{ m_0{\cal J}_T+m_2 {\cal J}_T }{2}\right)^{d/(2-d)},\nonumber\\
 m_2&=f(d)\left(\frac{m_1	 {\cal J}_T +m_3 {\cal J}_T }{2}\right)^{d/(2-d)},\nonumber\\
 m_3&=f(d)\left(\frac{m_2 {\cal J}_T +m_0  \delta {\cal J}}{2}\right)^{d/(2-d)},
\label{coupledeqs}
\end{align}
where $f(d)$ is a constant that only depends on the scaling dimension $d = \frac{1}{2 K_c}$.
The explicit expression for $f(d)$ is:
\begin{align}
 f(d)=&\frac{(1+\xi) \pi \Gamma(1-d/2)}{16 \sin \pi\xi\ \Gamma(d/2)}
\left(\frac{\Gamma(\frac{1}{2}+\frac{\xi}{2})\Gamma(1-\frac{\xi}{2})}
{4\sqrt{\pi}}\right)^{(d-2)}\nonumber\\
&\times\left( 2\sin \frac{\pi\xi}{2} \right)^d 
\left(\frac{\pi\Gamma(1-d/2)}{\Gamma(d/2)}\right)^{d/(2-d)}\nonumber\\
&\times\left(
\frac{2\Gamma(\xi/2)}{\sqrt{\pi}\Gamma(\frac{1}{2}+\frac{\xi}{2})}\right)^d
\label{constant}
\end{align}
Notice that the system of Eqs. \eqref{coupledeqs} is non-linear. Nevertheless it is easy to see that $m_0$ and
$m_3$ ($m_1$ and $m_2$) will take the same value ($m_0=m_3$ and $m_1=m_2$). We can therefore reduce Eq. \eqref{coupledeqs} to a system of 
only two coupled equations:
\begin{align}
 m_0&=f(d)\left(\frac{ m_0\delta {\cal J}+ m_1{\cal J}_T}{2}\right)^{d/(2-d)}\label{coupledeqs1}\\
 m_1&=f(d)\left(\frac{ m_0{\cal J}_T+m_1{\cal J}_T }{2}\right)^{d/(2-d)}
\label{coupledeqs2}
\end{align}
Taking the ratio of Eq. \eqref{coupledeqs1} and Eq. \eqref{coupledeqs2} we get:
\begin{equation}
 x=\left(\frac{\lambda x+1}{x+1}\right)^{d/(2-d)}
 \label{xeq}
\end{equation}
where $\lambda=\delta {\cal J}/{\cal J}_T$. 

We can solve numerically the previous transcendental Eq.\eqref{coupledeqs}, or directly solve the system 
Eqs. \eqref{coupledeqs1}-\eqref{coupledeqs2}. 
Before solving the system of equations \eqref{coupledeqs1}-\eqref{coupledeqs2} numerically for some values of the parameters, 
let us comment on Eq. \eqref{xeq}.

In the limiting case where ${\cal J}_T=\delta {\cal J}$ (i.e. ${\cal J}_{AA}^{\prime} =0 $) Eq. \eqref{xeq} has
the trivial solution $x=1$. In this case all the SC order parameters are in phase in the case $\delta {\cal J}>0$. On the other hand, 
for $\delta {\cal J}<0$, there is a shift of $\pi$ every four lattice spacings. So, in this case, the periodicity of the PDW order parameter 
will be eight (and not four), although the self-consistency equations actually will take the same form. 

For now we will assume $\delta {\cal J}>0$ (see section \ref{sec:P8PDW} for the $\delta {\cal J}<0$ case). 
Then in the pattern that we consider here, we find $x<1$ and so there is a coexistence between 
the uniform SC and the period 4 PDW order parameters. Let us now solve the system of equations \eqref{coupledeqs1}-\eqref{coupledeqs2} 
numerically for some values of the parameters. The results are summarize in table \ref{table:numerics}.

We now compute $T_c$ for this case. Following the same steps as in the previous section we have that:
\begin{align} 
H_0 &= H^{(0)}_0 - h_0 \int dx \cos(\sqrt{2\pi} \theta_0) \nonumber\\ 
H_1 & =H^{(0)}_1 - h_1 \int dx \cos(\sqrt{2\pi} \theta_1)
\end{align} 
and 
\begin{align} 
h_0 & =  \delta\cJ m_0+\cJ_T m_1\nonumber\\ 
h_1 & = \cJ_T m_0 + \cJ_T m_2 
\end{align} 
where we have used that $m_0=m_3$ and $m_1=m_2$. Since all the $A$-systems are equivalent,  they have the same SC susceptibility $\chi$. 
Then,  the self-consistency equations are 
\begin{equation} 
m_0  = \chi_{SC} h_0, \qquad 
m_1  = \chi_{SC} h_1
\end{equation} 
We can write this as a system of linear equations,
\begin{equation}
\left( \begin{array}{cc}
1-\chi_{SC}\delta\cJ & -\chi_{SC}\cJ_T \\
-\chi_{SC}\cJ_T & 1-\chi_{SC}\cJ_T\\
\end{array} \right)
\left( \begin{array}{c}
m_0\\
m_1\\
\end{array} \right)=
\left( \begin{array}{c}
0\\
0\\
\end{array} \right)
\end{equation}
which has a non-trivial solution if and only if  the determinant of the $2\times2$ matrix is zero. This gives us a quadratic equation in for $\chi_{SC}$. 
Choosing the positive solution we find that the critical temperature for the coexisting state is
\begin{equation}
 T_c=\Delta_s \left(\frac{2\cJ_T(\cJ_T-\delta\cJ)}{-\cJ_T-\delta\cJ+\sqrt{5\cJ_T^2-2\cJ_T\delta\cJ+\delta\cJ^2}} \right)^{\alpha}
\label{Tccoex}
 \end{equation}
where we recall that the exponent is given by  $\alpha=\displaystyle\frac{1}{2- 1/K_c}$. Notice that, in the limit $\delta\cJ\to\cJ_T$, we recover Eq. \eqref{Tcpure}. 
Thus,  as in the uniform or pure period 2 PDW state, $T_c$ has a power law behavior in
$\cJ_T$ and $\delta\cJ$, and it is not exponentially small as it would be in a weak coupling BCS type theory.



\begin{center}
\begin{table}[t]
\begin{tabular}{c c c c c c c}
\hline
\hline
{${\cal J}_T$} & {$\delta {\cal J}$} & {$d$} & {$m_0$} &{$m_1$} &{$\overline{m}$} & {$m_{\text{PDW}}$}\\
\hline
1 & 1   & 1/4 & 0.890893 & 0.890893 & 0.890893 & 0\\	
1 & 0.8 & 1/4 & 0.876601 & 0.889789 & 0.883195 & 0.0065943\\	
1 & 0.5 & 1/4 & 0.853007 & 0.887947 & 0.870477 & 0.0174703\\	
1 & 0   & 1/4 & 0.806035 & 0.884205 & 0.845120 & 0.0390853\\	
\hline
\hline
\end{tabular}
\caption{Numerical solution for the system of equations \eqref{coupledeqs2} for different values of the parameters 
${\cal J}_T$, $\delta {\cal J}$ and $d=1/2K_c$. 
We also define $\overline{m}=(m_1+m_0)/2$ and $m_{\text{PDW}}=(m_1-m_0)/2$, which correspond to the uniform and PDW part of the SC 
order parameter.}
\label{table:numerics}
\end{table}
\end{center}

\section{Fermionic Quasiparticles of the Superconducting States}
\label{sec:LLsystems}

So far, we have solved the coupled LE systems in the limit $|{\cal J}_{AA,j}| \gg |{\cal J}_{AB}|$ and $|{\cal J}_{AA,j}| \gg |{\cal J}_{AB}'|$ 
in Eq.\eqref{Hint} so that the couplings of the LE systems to eLL systems can be taken as the perturbation. 
In this limit, we have ignored the type-$B$ eLL systems and shown that the various SC states 
can emerge. Now we include the eLL systems back and investigate the nature of the full emergent SC state by looking at the SC proximity effect. 
First of all, we note that the eLL systems themselves will flow to the 2D Fermi liquid fixed point (at low enough temperatures) 
under the effect of the  hopping amplitude $t_{BB}$. This is the most 
relevant coupling in Eq.\eqref{Hint}. The result, for $t_{BB}$ small enough, is an anisotropic Fermi liquid with  an open Fermi surface, shown as the dashed curves in Fig.\ref{FSpurePDW}.

Having solved the largest energy scales in Eq.\eqref{Hint}, set by $t_{BB}$ and ${\cal J}_{AA}$, we now include the effect of the pair tunneling processes mixing the systems $A$ with the systems $B$, presented   
in Eq.\eqref{Hint}, and parametrized by the coupling constants ${\cal J}_{AB}$ and ${\cal J}_{AB}'$, respectively. We will study the effects of the SC states on the $A$ systems on the $B$ systems by treating the pair-tunneling terms to the lowest non-trivial order in perturbation theory in these coupling constants. Hence, we are assuming  that the interaction with the type-$B$ eLL systems does not back react to considerably
change the MFT value of the SC gap in the LE systems. 
As in Ref.[\onlinecite{granath-2001}], under the proximity effect mechanism the $B$ systems become superconducting and provide the  quasiparticles for the combined $A$-$B$ system. 

Since we are interested in the effect of the SC order parameters on the electronic spectrum, we replace the pair density $\Delta_{A, j} (x)$ of the type-$A$ LE systems in Eq.\eqref{Hint} by its MF value 
$\langle \Delta_{A,j} \rangle$ determined by the interchain MFT discussed in the previous Section \ref{sec:MFT}. 
In this approximation, we find that Eq.\eqref{Hint}  reduces to 
\begin{align}
&H^{\prime} \to \sum_{j}\int dx \Big\{
-t_{BB}\sum_{\sigma}[\psi_{B,j,\sigma}^{\dagger}\psi_{B,j+1,\sigma}+ {\rm h. c. }]\nonumber\\
&~-{\cal J}_{AB}[\Delta^{\dagger}_{B,j}\langle \Delta_{A,j}\rangle+\Delta^{\dagger}_{B,j}\langle \Delta_{A,j+1}\rangle +{\rm h. c.}]\nonumber \\
&~+{\cal J}_{AB}^{\prime}[\langle \Delta_{A,j}^{*}\rangle (\psi_{B,j,\uparrow}\psi_{B,j-1,\downarrow}+
\psi_{B,j-1,\uparrow}\psi_{B,j,\downarrow})+{\rm h. c.}]
\Big\}, 
\label{HPDWint2}
\end{align}
which is simply a theory of a Fermi surface coupled to the SC via a proximity coupling. Since Eq.\eqref{HPDWint2} is quadratic in the 
electron fields, we can readily diagonalize the effective Hamiltonian, and obtain the quasiparticle spectrum for the different   SC states 
found in Section \ref{sec:MFT}.

\subsection{Uniform SC phase and pure PDW phase} 

As we saw in section \ref{sec:staggered}, for the staggered (period 2) configurations of the Ising variables is 
possible to have  either a pure uniform SC state or a pure PDW state. The case of the uniform SC was studied by Granath, \etal \cite{granath-2001} 
who showed that, depending on the values of $\cJ_{AB}$ and $\cJ_{AB}^{\prime}$, it is possible  to have either a d-wave SC state with a fully gapped spectrum of 
quasiparticles or a  conventional d-wave SC state one with a nodal quasiparticle spectrum.  We refer the reader to their paper for 
further details.\cite{granath-2001} 

On the other hand, for the pure PDW state, even though the MF equation for the SC gap has the same form as for 
the uniform SC gap, the quasiparticle spectrum is quite different. We will study this spectrum in detail here. 
Let us start by defining the period 2 PDW order parameter, i.e. with ordering wave vector  ${\bm Q}=(0,\pi)$,
\begin{equation}
 \Delta^{A}_{j}=\Delta_{\bm Q} e^{i\pi j}
 \label{PDWOP}
\end{equation}
where $\Delta_{\bm Q}$ is given by the spin gap and the interchain MFT value for $\langle\cos\sqrt{2\pi}\theta\rangle$ 
which is given  in Section \ref{sec:unifstag} for the period 2 configuration of the Ising variables. Notice that for a period 2 state $\Delta_{\bm Q}=\Delta_{-\bm Q}$, since for a period 2 state $\bm Q$ and $-\bm Q$ differ by a reciprocal lattice vector.

\begin{figure}[hbt]
\centering
\subfigure[]{\includegraphics[width=0.23\textwidth]{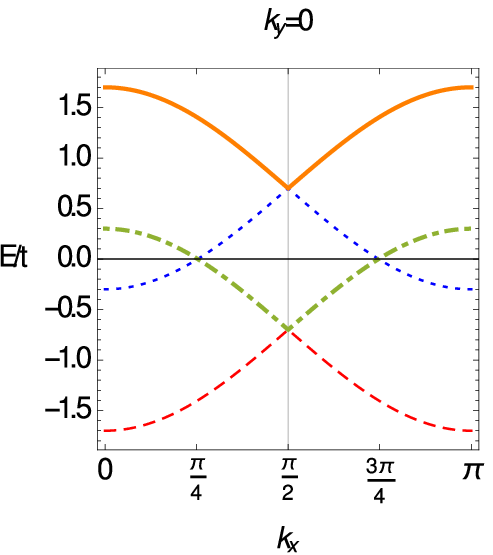} }
\subfigure[]{\includegraphics[width=0.23\textwidth]{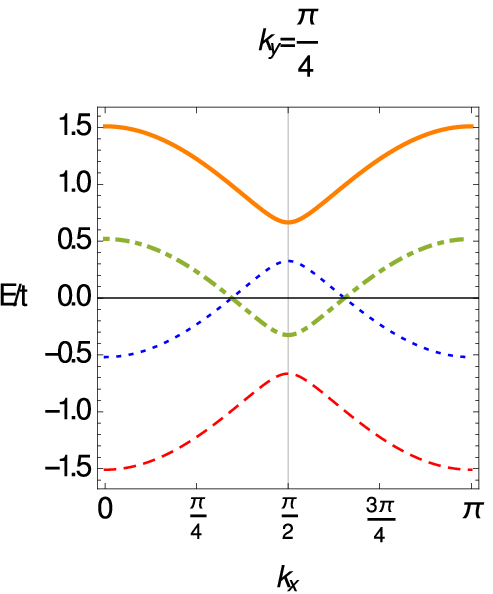}}
\subfigure[]{\includegraphics[width=0.23\textwidth]{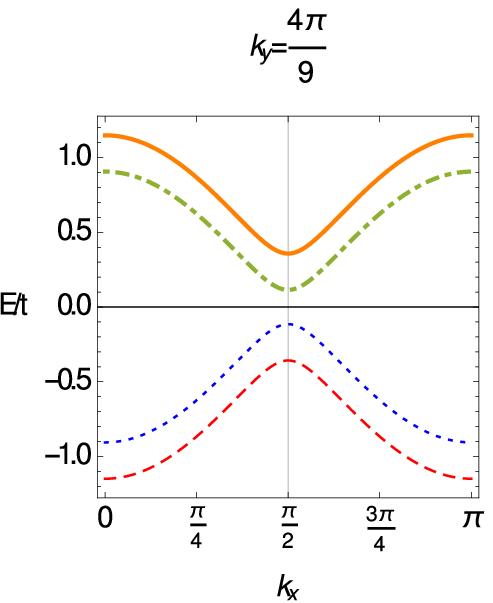}}
\subfigure[]{\includegraphics[width=0.23\textwidth]{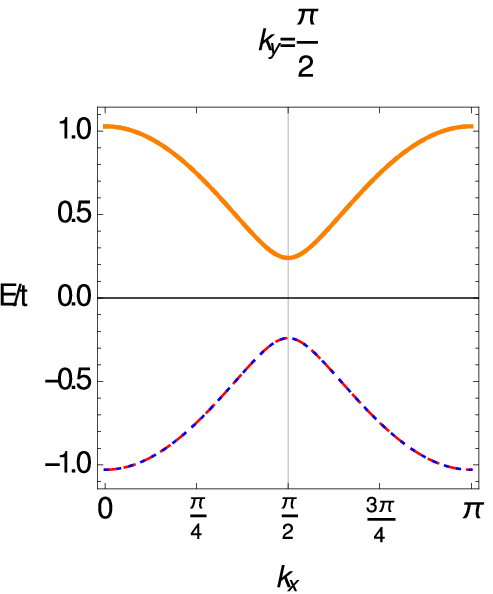}}
\caption{(Color online) a-d: Dispersion relation of the Bogoliubov quasiparticles for  $k_y=0, \frac{\pi}{4}, \frac{4\pi}{9}, \frac{\pi}{2}$, respectively,  for the period 2 PDW state. 
Here we used ${\cal J}_{AB}^{\prime}\Delta_{\bm Q}=0.12t$, $t_{BB}=0.7t$ and $\varepsilon(k_x)=-t\cos k_x$}
\label{bands}
\end{figure}

To find the quasiparticle spectrum we first
write down the Hamiltonian of Eq. \eqref{HPDWint2} in momentum space.
Defining the Nambu basis (here we  dropped the $B$ label in the electronic operators, since it is understood that we are referring to the eLL
systems) as:
\begin{equation}
 \Psi_\mathbf{k}^\dagger=(\psi_{\mathbf{k}\uparrow}^\dagger, \psi_{\mathbf{k}+(0,\pi)\uparrow}^\dagger,
\psi_{-\mathbf{k}\downarrow},\psi_{-\mathbf{k}-(0,\pi)\downarrow})
\label{NBPDW}
\end{equation}
we can write the Bogoliubov de-Gennes (BdG) Hamiltonian as 
\begin{equation}
H=\sum_{\mathbf{k}} \;\psi_\mathbf{k}^\dagger \;\hat{H}_\mathbf{k} \;\psi_\mathbf{k}
\end{equation}
where the one-particle Hamiltonian $\hat{H}_{\mathbf{k}}$ is given by
\begin{widetext}
\begin{equation}
\hat{H}_{{\bm k}}=\bmb 
 \varepsilon(k_x)-t_{BB}\cos(k_y ) & 0 & 0 & 2i{\cal J}_{AB}^{\prime}\Delta^*_{\bm Q}\sin(k_y) \\
 0 & \varepsilon(k_x)+t_{BB}\cos(k_y) &-2i{\cal J}_{AB}^{\prime}\Delta^*_{\bm Q}sin(k_y)& 0 \\
 0 & 2i{\cal J}_{AB}^{\prime}\Delta_{\bm Q}\sin(k_y) & -\varepsilon(k_x)+t_{BB}\cos(k_y) & 0 \\
 -2i{\cal J}_{AB}^{\prime}\Delta_{\bm Q}\sin(k_y) & 0 & 0 & -\varepsilon(k_x)-t_{BB}\cos(k_y)
\emb
\label{HBdGPDW}
\end{equation} 
\end{widetext}
From this one-particle Hamiltonian we find the quasiparticle spectrum 
\begin{equation}
 E(\mathbf{k})=\pm t_{BB} \cos(k_y)\pm \sqrt{\varepsilon^2(k_x)+4{\cal J}_{AB}^{\prime 2}|\Delta_{\bm Q}|^2\sin^2(k_y)}
\end{equation}
\begin{figure}[hbt]
\centering
\hbox{\includegraphics[width=0.46\columnwidth]{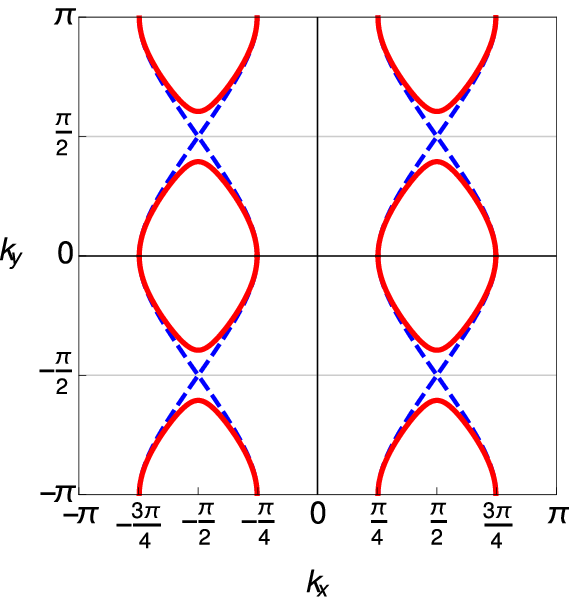} 
\hskip -.1cm 
\includegraphics[width=0.55\columnwidth]{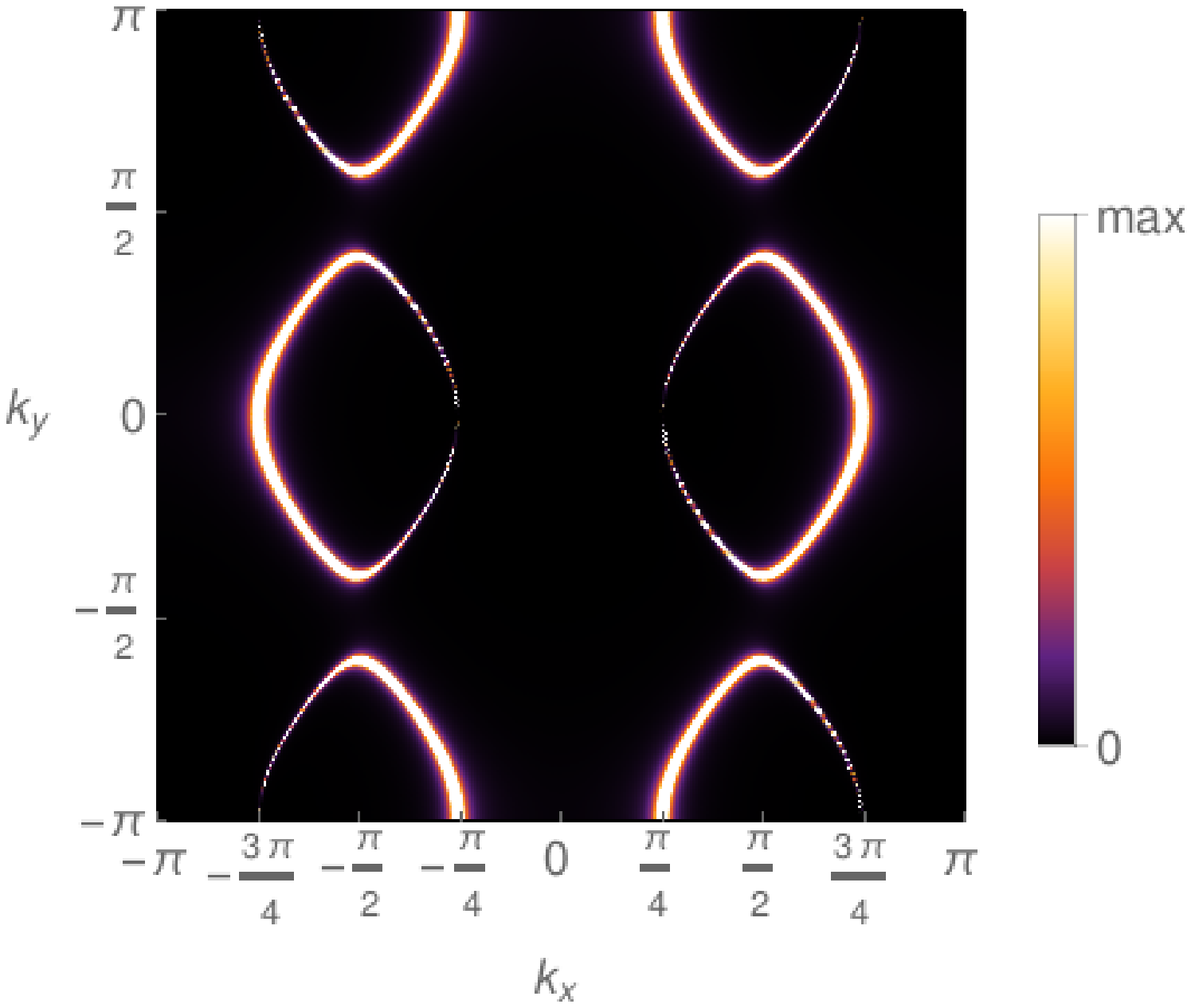}}
\caption{(Color online) On the left, Fermi surface for the pure period 2 PDW state. The dashed (blue) line corresponds to the original
FS in the absence of superconductivity. The solid (red) line corresponds to the new FS after the superconducting proximity state is established.
On the right, the spectral function $A(\mathbf{k},0)$ corresponding to the pockets on the left.
We used ${\cal J}_{AB}^{\prime}\Delta_{\bm Q}=0.12t$, $t_{BB}=0.7t$, $\varepsilon(k_x)=-t\cos k_x$ and $\delta=10^{-4}t$}
\label{FSpurePDW}
\end{figure}
In Fig. \ref{FSpurePDW} we plot the Fermi surface of the Bogoliubov quasiparticles of this period 2 PDW state for some values of the parameters. 
In contrast to the pure uniform SC state, whose  spectrum can be either nodal or fully gapped, we 
find that   this PDW state ($\Delta_{\bm Q}\neq0$ in Eq. \eqref{PDWOP}) has pockets of Bogoliubov quasiparticles, 
as it is also found in the weak coupling theories.\cite{baruch-2008,Berg-2009,Loder-2010,Radzihovsky-2011,zelli-2012,lee-2014} The size of the pockets depends
on the strength of the SC gap.
In addition, we compute the spectral function given by (see for instance Ref. [\onlinecite{Seo-2008}]):
\begin{equation}
 A(\mathbf{k},\omega)=-\frac{1}{\pi}\text{Im}[\hat{G}_{11}(\mathbf{k},\omega)]
 \label{spectralfn}
\end{equation}
where 
\begin{equation}
\hat{G}(\mathbf{k},\omega)=\frac{1}{\omega+i\delta-\hat{H}_{{\bm k}}}
\end{equation}
 is the retarded Green function and $\delta=0^+$. 
The spectral function $A(\mathbf{k},\omega=0)$ for this pure period 2 PDW state is shown in Fig. \ref{FSpurePDW}.
In Fig. \ref{bands} we plot the dispersion relation of the Bogoliubov excitations for several values of  $k_y$.

\subsection{Coexistence Phase of a Period 4 PDW and a uniform SC: the Striped Superconductor}

We start by writing the SC order parameter, which includes both the uniform SC and the PDW order parameters as an expansion of the form
\begin{equation}
 \Delta^{A}_{j}=\Delta_0+\sqrt{2}\Delta_{\bm Q}\cos\left(\frac{\pi j}{2}+\frac{\pi}{4}\right)
 \label{coexOP}
\end{equation}
where the expectation values of the order parameters $\Delta_0$ and $\Delta_{\bm Q}$, where ${\bm Q}=(0,\frac{\pi}{2})$ is the ordering wave vector,  are set jointly by the spin gap of the LE systems and by the interchain MFT value for $\langle\cos\sqrt{2\pi}\theta\rangle$ 
found in the previous section for the period 4 state of the Ising degrees of freedom.  

\begin{figure}[t]
\centering
\includegraphics[width=0.78\columnwidth]{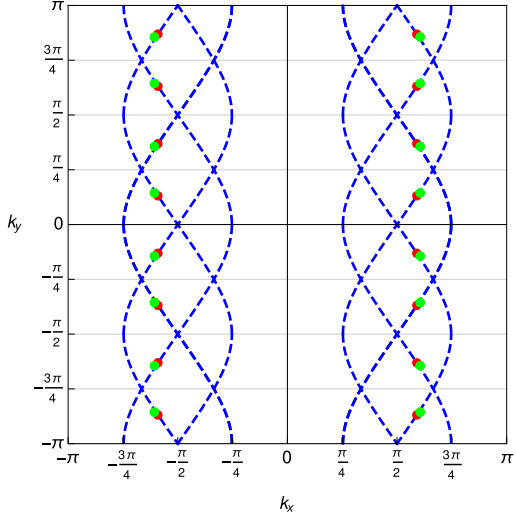}
\caption{
(Color online) Quasiparticle spectra with nodal points in the coexistence phase. 
The dashed (blue) line corresponds to the original FS in the absence of superconductivity. 
The red points correspond to the position of the nodes in the absence of the PDW state ($\Delta_{\bm Q}=0$).
The green points correspond to the position of the nodes in the presence of the PDW state with $\Delta_{\bm Q}=0.2$.
We have chosen the parameters ${\cal J}_{AB}^{\prime}=0.5t$, 
${\cal J}_{AB}=0.2t$, $t_{BB}=0.7t$, $\Delta_0=0.2$ and $\varepsilon(k_x)=-t\cos k_x$}.
\label{nodes}
\end{figure}
We now write down the Hamiltonian in momentum space  following the notation of Ref. [\onlinecite{baruch-2008}].
We define the Nambu spinor as:
\begin{equation}
 \Psi_\mathbf{k}^\dagger=(\psi_{\mathbf{k}\uparrow}^\dagger, \psi_{\mathbf{k}+\mathbf{q}\uparrow}^\dagger,
\ldots,\psi_{-\mathbf{k}\downarrow},\psi_{-(\mathbf{k}+\mathbf{q})\downarrow}, \ldots)
\label{NB}
\end{equation}
where $\mathbf{q}$ is the ordering wavevector. In our case $\mathbf{q}=(0,\pi/2)$ and $\mathbf{k}$ is taking values over the reduced Brillouin 
zone (RBZ) associated with the ordered state, which in this case is $k_x \in [-\pi, \pi)$ and  $k_y\in [-\pi/4, \pi/4)$. 
In this basis the Hamiltonian is given by:
\begin{equation}
H=\sum_{\mathbf{k}\in RBZ} \;\psi_\mathbf{k}^\dagger \;\hat{H}_\mathbf{k} \;\psi_\mathbf{k},
\label{Hrep}
\end{equation}
where  the BdG Hamiltonian $\hat{H}_{\bm k}$ in the Nambu basis of Eq.\eqref{NB} is given by:
\begin{equation}
 \hat{H}_\mathbf{k}=\left( \begin{array}{cc}
\mathcal{A}_\mathbf{k} & \mathcal{C}_\mathbf{k} \\
   \mathcal{C}^\dagger_\mathbf{k} & -\mathcal{A}_\mathbf{k}
\end{array} \right)
\label{Hmatrix}
\end{equation}
where $\mathcal{A}_\mathbf{k}=\textrm{diag} (\varepsilon(\mathbf{k}),
\varepsilon(\mathbf{k}+\mathbf{q}),\ldots)$ is a diagonal matrix, and the square matrix $\mathcal{C}_\mathbf{k}$ contains the SC order parameters.
Since the ordering vector is $\pi/2$ along the $k_y$ direction, our matrix $\mathcal{C}_\mathbf{k}$ is given by a $4\times 4$ 
matrix with the form:
\begin{equation}
 \!\!\mathcal{C}_\mathbf{k}=\left( \begin{array}{cccc}
f_0(\mathbf{k}) & f_1(\mathbf{k})  & f_2(\mathbf{k})  & f_3(\mathbf{k}) \\
 f_1^*(\mathbf{k}) & f_0(\mathbf{k}+\mathbf{q}) &  f_1(\mathbf{k}+\mathbf{q}) &  f_2(\mathbf{k}+\mathbf{q}) \\
 f_2^*(\mathbf{k})  & f_1^*(\mathbf{k}+\mathbf{q})  &  f_0(\mathbf{k}+2\mathbf{q})  & f_1(\mathbf{k}+2\mathbf{q})\\
  f_3^*(\mathbf{k})  & f_2^*(\mathbf{k}+\mathbf{q})  & f_1^*(\mathbf{k}+2\mathbf{q}) & f_0(\mathbf{k}+3\mathbf{q})  \\
\end{array} \right)
\label{Cmatrix}
\end{equation}
where $f_0$ corresponds to uniform pairing and $f_1,f_2,f_3$ to the finite momentum pairing. The explicit expressions are the following:
\begin{align}
 f_0(\mathbf{k})&=2\Delta_0({\cal J}_{AB}-{\cal J}_{AB}^{\prime}\cos k_y)\nonumber\\
 f_1(\mathbf{k})&=-i\Delta_{\bm Q}({\cal J}_{AB}-\sqrt{2}{\cal J}_{AB}^{\prime}\cos(k_y+q_y/2))\nonumber\\
 f_2(\mathbf{k})&=0\nonumber\\
 f_3(\mathbf{k})&=i\Delta_{\bm Q}({\cal J}_{AB}-\sqrt{2}{\cal J}_{AB}^{\prime}\cos(k_y-q_y/2))
\end{align}
where we recall that $\mathbf{q}=(0,\pi/2)$, so $q_y=\pi/2$.
\begin{figure}[t!]
\centering
\subfigure[]{\includegraphics[width=0.28\textwidth]{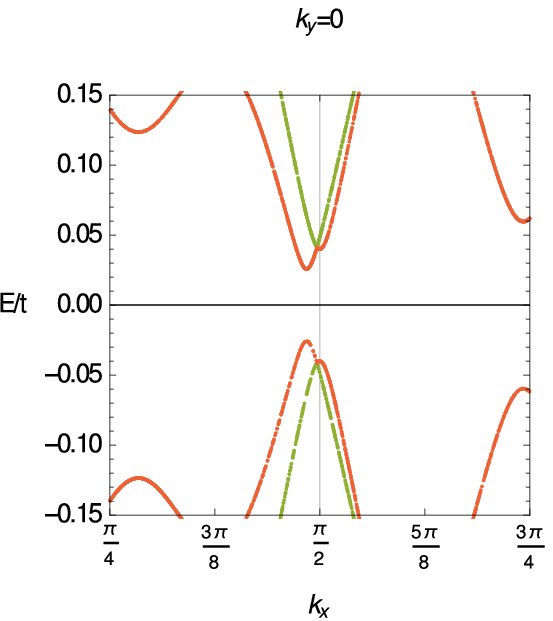} }
\subfigure[]{\includegraphics[width=0.28\textwidth]{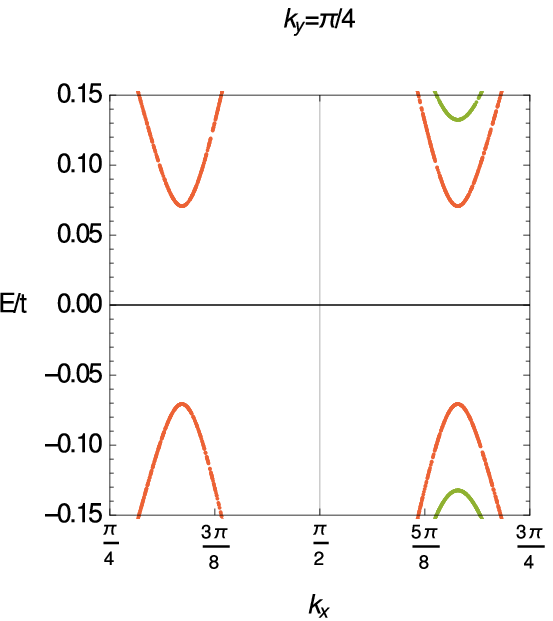}}
\subfigure[]{\includegraphics[width=0.28\textwidth]{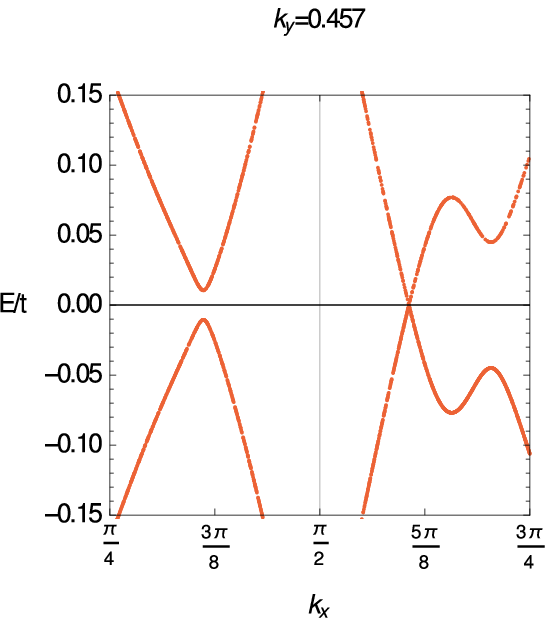}}
\caption{(Color online) Dispersion relations of the quasiparticles in the coexistence phase shown for several values of $k_y$. Notice that the dispersion relation is only gapless for $k_y\approx0.457$, which
corresponds to the position of the nodal point for the same set of parameters used in Fig. \ref{nodes}. ${\cal J}_{AB}^{\prime}=0.5t$, 
${\cal J}_{AB}=0.2t$, $t_{BB}=0.7t$, $\Delta_0=\Delta_{\bm Q}=0.2$ and $\varepsilon(k_x)=-t\cos k_x$}
\label{bandscoex}
\end{figure}

First of all, 
due to the periodicity of the PDW SC state, it is necessary to the fold original FS.
Let us first analyze the case of the pure uniform SC state. In this case ($\Delta_{\bm Q}=0$) the spectrum can be easily calculated
from the Hamiltonian given in eq. \eqref{Hmatrix}:
\begin{align}
E_{1,\pm}^{2}&=(\varepsilon(k_x)\pm t_{BB}\cos(k_y))^2+\Delta_0^2(\cJ_{AB}\pm\cJ_{AB}^{\prime}\cos(k_y))^2\nonumber \\
E_{2,\pm}^{2}&=(\varepsilon(k_x)\pm t_{BB}\sin(k_y))^2+\Delta_0^2(\cJ_{AB}\pm\cJ_{AB}^{\prime}\sin(k_y))^2
\end{align}
We can see that this SC state will have a quasiparticle spectrum with  nodes if $|\cJ_{AB}|<|\cJ_{AB}^{\prime}|$. Now, even in the coexistence phase, where both $\Delta_{\bm Q}\neq0$ 
and $\Delta_0\neq0$, the quasiparticle spectrum may still can have nodes. For the pure uniform SC state, the position of the nodes depends on
the values $\cJ_{AB}/\cJ_{AB}^{\prime}$ and $t_{BB}$. In the coexistence phase the position of the nodes will depends on $\Delta_{\bm Q}$ as well
(see Fig. \ref{nodes}).
As in the case of pure period 2 PDW state, we show in Fig. \ref{bandscoex} the dispersion relation of the quasiparticles for several  values of $k_y$.

\subsection{Period 8 PDW state}
\label{sec:P8PDW}

\begin{figure}[hbt]
\centering
\subfigure[]{\includegraphics[width=0.4\textwidth]{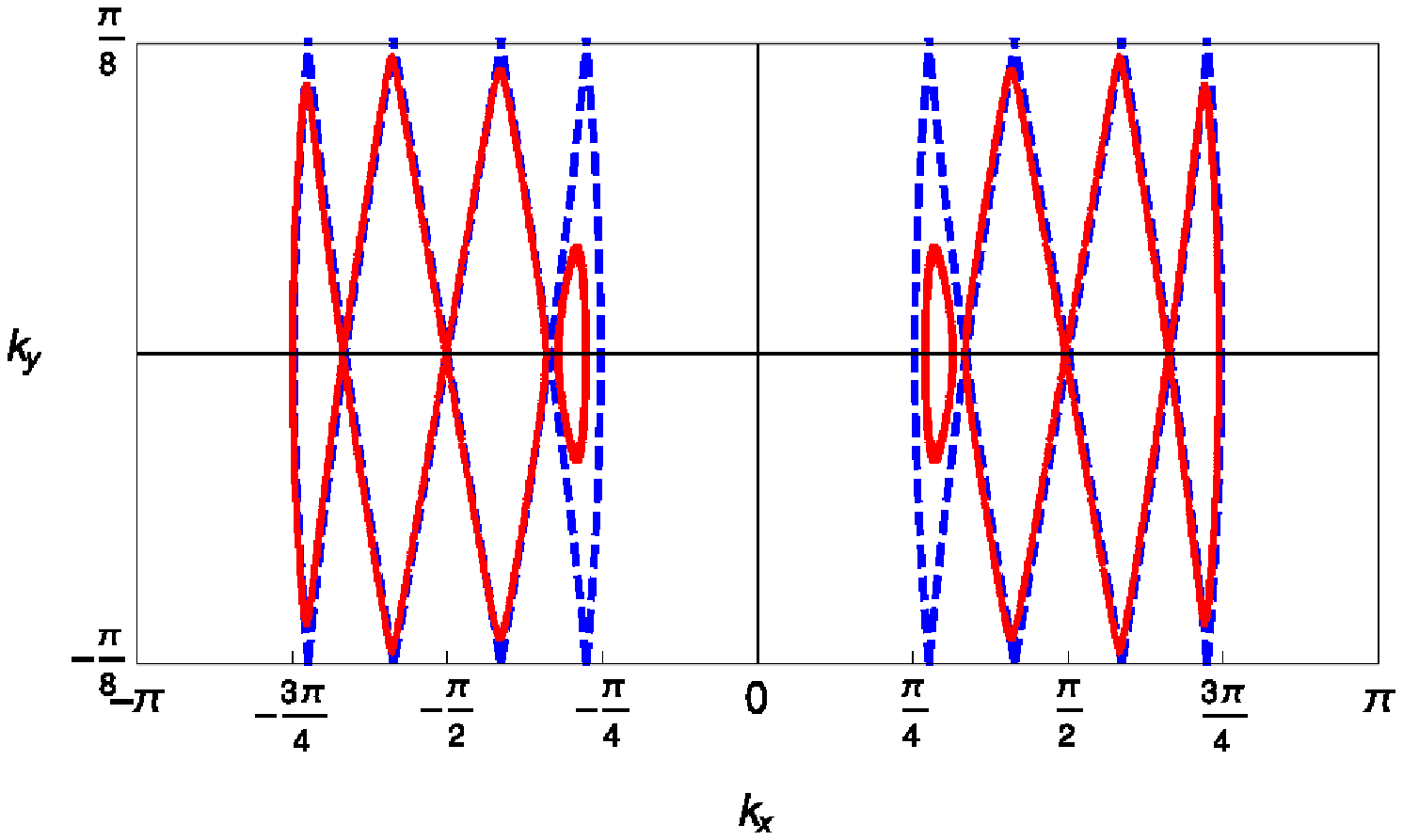} }
\subfigure[]{\includegraphics[width=0.4\textwidth]{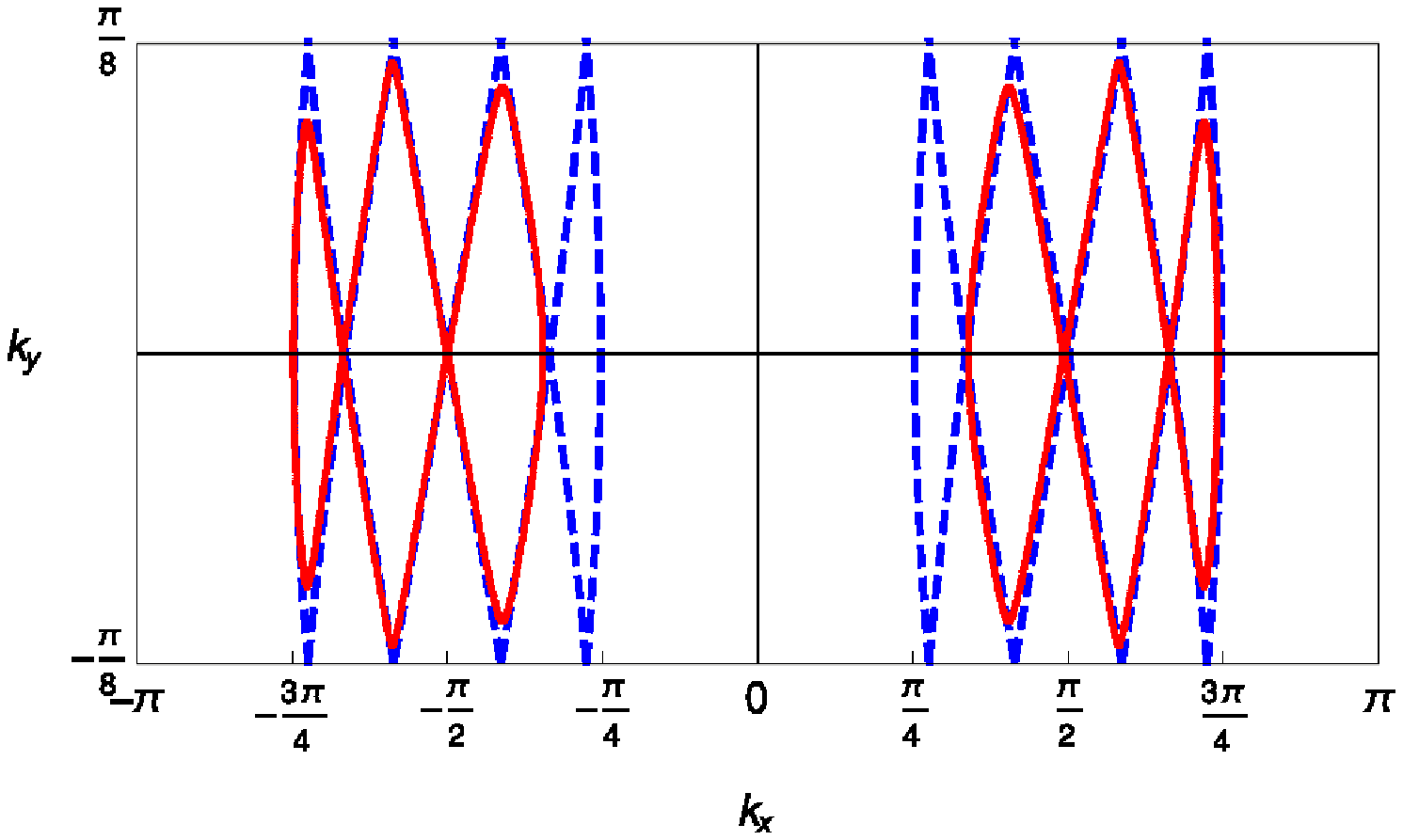}}
\subfigure[]{\includegraphics[width=0.4\textwidth]{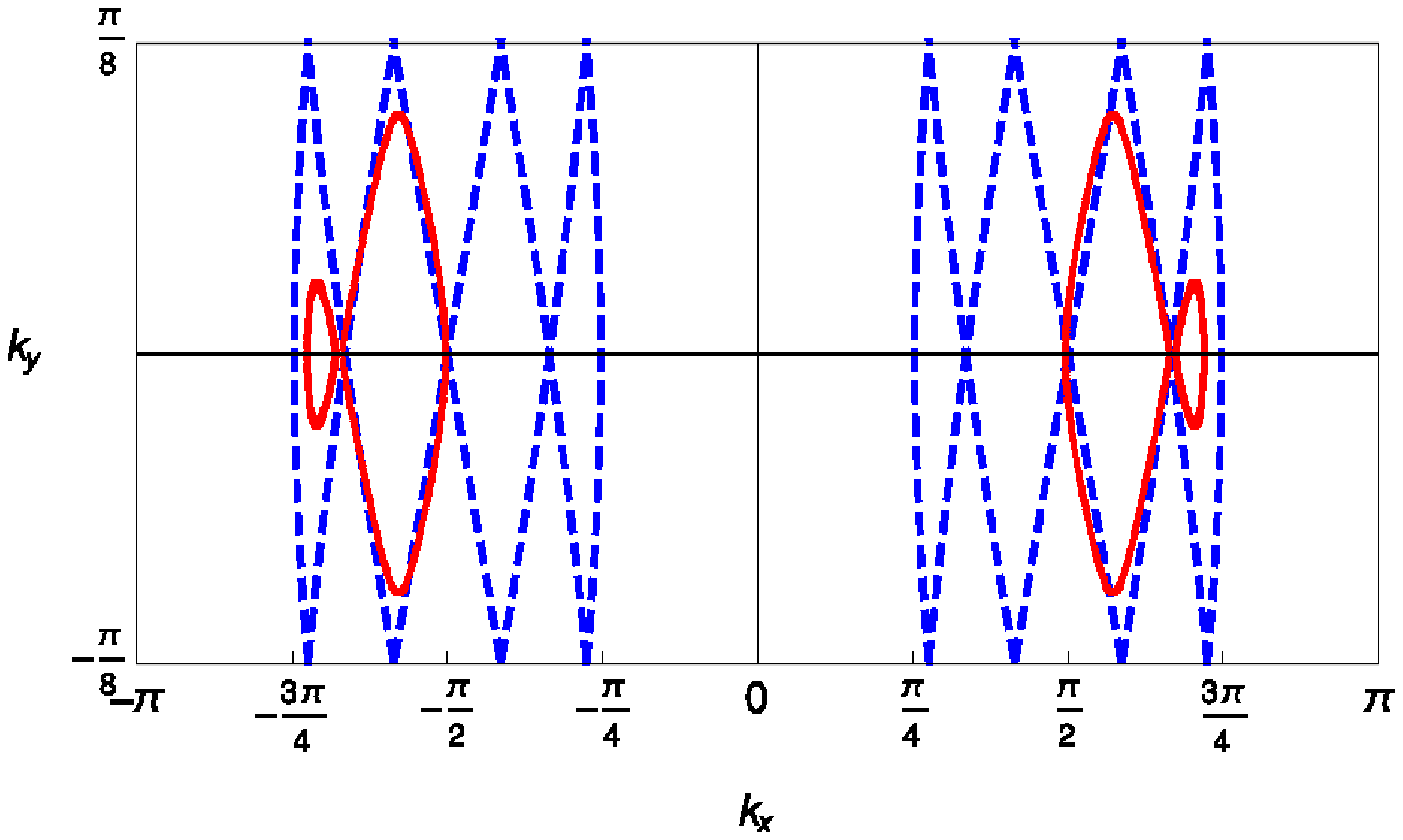}}
\caption{(Color online) FS for the period 8 PDW state. 
The dashed (blue) line corresponds to the original FS in the absence of superconductivity. 
The solid (red) line corresponds to the new FS (pockets) after the introduction of superconductivity ($\Delta_1\neq0$ and $\Delta_2\neq0$).
In (a) $\Delta_1=\Delta_2=0.05$, in (b) $\Delta_1=0.08$ and $\Delta_2=0.1$ and in (c) $\Delta_1=0.25$ and $\Delta_2=0.3$
In all figures ${\cal J}_{AB}^{\prime}=0.6t$, ${\cal J}_{AB}=0.4t$, $t_{BB}=0.7t$, $\Delta_0=0$ and $\varepsilon(k_x)=-t\cos k_x$}
\label{pocketsP8}
\end{figure}

Above we focused on the coexistence phase for the period 4 case. This was the case when $\delta\cJ>0$. However, 
if $\delta {\cal J}<0$ (i.e. for $\mathcal{J}_{AA}<\mathcal{J}'_{AA}$) the case is different and we find a PDW state. There is a shift of $\pi$ every four lattice sizes, 
so in this case the periodicity of the PDW order parameter is actually eight (not four!).
Nevertheless, the self-consistency equations will have the same form:
\begin{align}
 m_0&=f(d)\left(\frac{ m_0|\delta {\cal J}|+ m_1{\cal J}_T}{2}\right)^{d/(2-d)}\nonumber\\
 m_1&=f(d)\left(\frac{ m_0{\cal J}_T+m_1{\cal J}_T }{2}\right)^{d/(2-d)}
\label{coupledeqsP8}
\end{align}
The pattern of the SC order parameter is now that of a pure period 8 PDW SC state:
$$\Delta=(\Delta_1,\Delta_2,\Delta_2,\Delta_1,-\Delta_1,-\Delta_2,-\Delta_2,-\Delta_1,\Delta_1,\ldots)$$
We can write the previous pattern using the following SC order parameter:
\begin{equation}
 \Delta^{A}_{j}=\Delta\sin\left(\frac{\pi j}{4}+\frac{\pi}{8}\right)+\tilde{\Delta}\sin\left(\frac{3\pi j}{4}+\frac{3\pi}{8}\right)
 \label{Period8OP}
\end{equation}
where we have defined:
\begin{align}
 \Delta&=\Delta_1\sin\left(\frac{\pi}{8}\right)+\Delta_2\cos\left(\frac{\pi}{8}\right)\nonumber\\
 \tilde{\Delta}&=\Delta_1\cos\left(\frac{\pi}{8}\right)-\Delta_2\sin\left(\frac{\pi}{8}\right)
 \end{align}
where $\Delta_1$ and $\Delta_2$ are given by the spin gap and the interchain MFT value for $\langle\cos\sqrt{2\pi}\theta\rangle$ 
in eq. \eqref{coupledeqsP8}.

Since we are dealing with a period 8 SC state, the reduced Brillouin Zone is now
case is $k_x \in [-\pi, \pi)$ and  $k_y\in [-\pi/8, \pi/8)$ and $\mathbf{q}=(0,\pi/4)$. The difference between the period 4 and the
period 8 is that the definition of the $\mathcal{C}_{\mathbf{k}}$ matrix is different, since is now an $8\times8$ matrix.

\begin{equation}
 \!\!\mathcal{C}_\mathbf{k}=\left( \begin{array}{cccc}
f_0(\mathbf{k}) & f_1(\mathbf{k})  & \cdots  & f_7(\mathbf{k}) \\
 f_1^*(\mathbf{k}) & f_0(\mathbf{k}+\mathbf{q}) &  \cdots &  f_6(\mathbf{k}+\mathbf{q}) \\
 \vdots &  & \ddots \\
  f_7^*(\mathbf{k})  &  &  & f_0(\mathbf{k}+7\mathbf{q})  \\
\end{array} \right)
\label{CmatrixP8}
\end{equation}
where the $f_i(\mathbf{k})$'s are given by the following expressions:
\begin{align}
 f_0(\mathbf{k})=&f_2(\mathbf{k})=f_4(\mathbf{k})=f_6(\mathbf{k})=0\nonumber\\
 f_1(\mathbf{k})=&i\Delta\left(\frac{1}{2}{\cal J}_{AB}(e^{-i\pi/8}+e^{-i3\pi/8})\right.\nonumber\\
 &\qquad\left.-{\cal J}_{AB}^{\prime}e^{-i\pi/4}\cos(k_y+q_y/2)\right)\nonumber\\
 f_3(\mathbf{k})=&i\tilde{\Delta}\left(\frac{1}{2}{\cal J}_{AB}(e^{-3i\pi/8}+e^{-i9\pi/8})\right.\nonumber\\
 &\qquad\left.-{\cal J}_{AB}^{\prime}e^{-i3\pi/4}\cos(k_y+3q_y/2)\right)\nonumber\\
 f_5(\mathbf{k})=&-i\tilde{\Delta}\left(\frac{1}{2}{\cal J}_{AB}(e^{3i\pi/8}+e^{i9\pi/8})\right.\nonumber\\
 &\qquad\left.-{\cal J}_{AB}^{\prime}e^{i3\pi/4}\cos(k_y-3q_y/2)\right)\nonumber\\
 f_7(\mathbf{k})=&-i\Delta\left(\frac{1}{2}{\cal J}_{AB}(e^{i\pi/8}+e^{i3\pi/8})\right.\nonumber\\
 &\qquad\left.-{\cal J}_{AB}^{\prime}e^{i\pi/4}\cos(k_y-q_y/2)\right)\nonumber
 \end{align}
where we recall that $\mathbf{q}=(0,\pi/4)$, so $q_y=\pi/4$.
Having $\mathcal{C}_{\mathbf{k}}$ we can write down our BdG Hamiltonian as in eq. \eqref{Hmatrix}.
In Fig. \eqref{pocketsP8} we show the FS for some values of $\Delta_1$ and $\Delta_2$. As in the pure period 2 PDW state, we see the formation
of pockets due to the folding of the FS.

\section{Other phases} 
\label{sec:phases}

For completeness we summarize the other possible phases occurring in the system.  Following closely Granath \etal \cite{granath-2001} 
we treat the interactions appearing in eq. \eqref{Hint} perturbatively around the so called decoupled fixed point. At this
fixed point (FP) the systems are completely decoupled, and each one of the systems corresponds to a 1D system 
that can be solved using bosonization. 
Around the decoupled FP a perturbation with coupling constant $g$ is relevant (irrelevant) if its scaling dimension $d_g<2$ ($d_g>2$). 
The scaling dimensions for the operators appearing in Eq. \eqref{Hint} are given in the work of Granath {\it et al.}.\cite{granath-2001}
The phases found by Granath {\it et al.} are:
\begin{enumerate}
 \item 
Typically, the couplings between the eLL and LE systems are irrelevant or less relevant than the coupling between $AA$ and $BB$ 
 systems separately. In this case the RG  flows to the point where all the $AB$ couplings go to zero. At this FP the system is made of 
 two (independent) interpenetrating systems, $A$ and $B$. 
 \item 
 The $\cJ_{AA}$ ($\cJ_{AA}^{\prime}$) term is relevant for $K_{c}^{(A)}>1/2$. In this case the $A$ systems develop long-range order and a 
 full spin gap.
Since the $BB$ electron tunneling operator  has lower scaling dimension than the $BB$ spin exchange interaction, in the absence of a charge gap in 
 the $B$ subsystem, most probably the $B$ subsystem is in a anisotropic Fermi liquid phase. However, this two fluid FP is unstable due to
 the proximity effect. Depending on the parameters in the Hamiltonian of Eq.\eqref{Hint} the quasiparticle spectrum can be gapless 
 (present nodes or pockets in the pure PDW state) or fully gapped. This means that we can have several possible stable SC phases, 
 a SC state with Fermi pockets, a nodal SC state, or a fully gapped SC state. 
 These  were the phases studied in the previous sections using interchain MFT and coupling the eLL 
 systems to the LE systems.
 \item 
 If $\Delta_c^{(B)}>0$, the $B$ subsystem can develop a antiferromagnetic phase. At this FP will be a coexistence between 
 superconductivity (in the $A$ subsystem) and antiferromagnetism (in the $B$ subsystem). This FP is stable, due to the spin gap in the 
 SC ($A$) and the charge gap in the antiferromagnet ($B$). The quasiparticle spectrum is therefore fully gapped as is also found in BCS-type theories.\cite{Loder-2011}
 \end{enumerate}

\section{Concluding Remarks}
\label{sec:conclusions}

We have investigated a model of an array of two inequivalent systems in the quasi-one dimensional limit. 
In this limit we have treated the   interactions between the different systems in the array exactly using bosonization methods and   interchain mean field
theory. The phases that we found are either a uniform d-wave superconductor, a striped superconductor (in which the uniform SC and the PDW SC state coexist), and a PDW state. To simplify the analysis we only looked at the case in which the modulation of the SC state is commensurate. 

The resulting critical temperatures are, as expected, upper bounds on the actual physical critical temperatures. As emphasized in Refs.[\onlinecite{Arrigoni-2004}] and [\onlinecite{kivelson-2007}], the analytic dependence of these mean field $T_c$'s on the coupling constants obeys the exact power-law scaling behavior predicted by a renormalization group analysis of the dimensional crossover from the 1D regime to the full (but anisotropic) 2D phases, albeit with an overestimate of the prefactor. 

On the other hand, the actual critical temperatures are significantly suppressed from the values quoted here due to the the two-dimensional nature of the array. Hence we expect the ground states that we found here to undergo a sequence of thermodynamic phase transitions leading to a complex phase diagram of the type discussed by Berg {\it et al.}\cite{Berg-2009b} (and by Agterberg and Tsunetsugu.\cite{agterberg-2008}) It is well known from classical critical phenomena of 2D commensurate systems that states of the type we discuss here may become incommensurate at finite temperatures due to thermal fluctuations if the period of the ordered state is longer than a critical value (typically equal to four), see, e.g. Ref.[\onlinecite{Chaikin-1995}].

We have shown that a high energy scales (of the order of the spin gap), we can first determine the SC phases of one set of systems 
(in our notation, the Luther-Emery liquid systems $A$). At these energy scales we 
showed that it is possible to have, in addition to a uniform SC phase, a pure PDW state and a coexistence phase of a uniform and a PDW state. 
Having determined the SC in the LE systems, we proceeded to incorporate the electronic Luttinger liquid systems perturbatively.
We found that the quasiparticle spectrum arising from the eLL systems can present Fermi pockets if the SC state is a pure PDW state. 
In the case of coexistence uniform SC and PDW state or pure uniform SC (i.e. a striped superconductor) the quasiparticle spectrum can have nodes or be fully gapped depending
on the value of the coupling in the model.  
We should stress, as it was done recently in Ref. [\onlinecite{fradkin-2014}], that in this quasi-1D approach the superconducting state 
evolves from a local high energy scale, the spin gap, which hence has magnetic origin. For temperature $T$ higher that the spin  gap, the system 
is a quasi 1D system which does not have quasiparticles in the spectrum up to a scale, determined by an electron tunneling scale, to a crossover 
to a Fermi liquid type system. Hence, at least qualitatively, systems of this type behave as `high $T_c$ superconductors.'

\begin{acknowledgments}
We thank  Steven Kivelson for great discussions and V. Chua for his help generating the density plot for the spectral function.
This work was supported in part by the NSF grants DMR-1064319 (GYC,EF) and DMR 1408713 (EF) at the University of Illinois, 
DOE Award No. DE-SC0012368  (RSG) and  Program Becas Chile (CONICYT) (RSG).  
\end{acknowledgments}
 

\end{document}